\documentclass[%
 reprint,
 superscriptaddress,
 numerical,
 showpacs,
 amsmath,amssymb,
 aps,
 longbibliography,
 prd,
 floatfix
]{revtex4-2}

\usepackage{graphicx}

\usepackage[%
  colorlinks=true,
  urlcolor=blue,
  linkcolor=blue,
  citecolor=blue
]{hyperref}

\usepackage{braket}

\usepackage{lipsum}

\usepackage[normalem]{ulem}

\let\Re\relax
\let\Im\relax

\DeclareMathOperator{\tr}{tr}

\DeclareMathOperator{\Re}{Re}
\DeclareMathOperator{\Im}{Im}
\DeclareMathOperator{\Ei}{Ei}

\newcommand{\Creg}{C^{\mathrm{(reg)}}}

\def\tcm{T.C.M. Group, Cavendish Laboratory, University of Cambridge, J.J. Thomson Avenue, Cambridge, CB3 0HE, UK}
\def\tauni{Raymond and Beverly Sackler School of Physics and Astronomy, Tel Aviv University, Tel Aviv 6997801, Israel}

\begin{document}

\title{Quantum geometry of non-Hermitian systems}

\author{Jan Behrends}
\affiliation{\tcm}

\author{Roni Ilan}
\affiliation{\tauni}

\author{Moshe Goldstein}
\affiliation{\tauni}

\begin{abstract}
The Berry curvature characterizes one aspect of the geometry of quantum states.
It materializes, among other consequences, as an anomalous velocity of wave packets.
In non-Hermitian systems, wave packet dynamics is enriched by additional terms that can be expressed as generalizations of the Berry connection to non-orthogonal eigenstates.
Here, we contextualize these anomalous non-Hermitian contributions by showing that they directly arise from the geometry of the underlying quantum states as corrections to the distance between left and perturbed right eigenstates.
By calculating the electric susceptibility for a single-band wave packet and comparing it with the wave packet's localization, we demonstrate that these terms can, in some circumstances, lead to a violation of fluctuation-dissipation relations in non-Hermitian systems.
We discuss experimental signatures in terms of response functions and transport signatures.
\end{abstract}

\maketitle

The geometry of quantum states~\cite{Provost:1980hs,Berry:1989es,Torma:2023jw} plays a key role in modern condensed matter physics.
The Berry phase~\cite{Berry:1984ka}, acquired during cyclic adiabatic evolution and intrinsically tied to, e.g., polarization, orbital magnetism, and charge pumping~\cite{Xiao:2010kw}, equals, for infinitesimal loops, the Berry curvature~\cite{Heslot:1985bw,Ashtekar:1999iy,Facchi:2010hl}.
The Berry curvature can directly be measured in wave packet dynamics~\cite{Chang:1995gk,Chang:1996gv}, and its integral over closed manifolds is the quantized Chern number~\cite{Thouless:1982kq,Thouless:1983hb}.

A key object for the geometric formulation of quantum mechanics~\cite{Ashtekar:1999iy} is the quantum geometric tensor~\cite{Provost:1980hs,Berry:1989es}.
Its imaginary party equals (half) the Berry curvature, and its real part defines a metric of the projective Hilbert space~\cite{Study:1905ku}, the so-called Fubini-Study metric~\footnote{Naming conventions of these terms vary~\cite{Kolodrubetz:2017jg}. Throughout this work, we refer to the complex object $q_{\mu\nu}$ as the quantum geometric tensor, its real part the Fubini-Study metric~\cite{Study:1905ku,Provost:1980hs,Anandan:1990dx} (often called the quantum metric~\cite{Neupert:2013eu,Kolodrubetz:2013bf}), and twice its imaginary part the Berry curvature.}.
The Fubini-Study metric contributes to nonequilibrium dynamics~\cite{Anandan:1990dx,DeGrandi:2011kq}, quantum phase transitions~\cite{Yang:2008cl,Garnerone:2009ho}, localization of wave functions~\cite{Kudinov1991,Marzari:1997co,Resta:1999kl} current noise~\cite{Neupert:2013eu}, bounds on energy gaps~\cite{Kivelson:1982fr,Onishi:2024gl,Onishi2024} and superfluid weight~\cite{Peotta:2015gb,Torma:2023jw}, and it gives higher-order corrections to semiclassical equations of motion~\cite{Gao:2014cn}.

The Fubini-Study metric naturally arises from the fidelity of quantum states $F [\phi,\psi]$, which for pure states $\ket{\phi}$ and $\ket{\psi}$ is their squared overlap~\cite{Jozsa:1994ik}.
The angle $\gamma = \arccos \sqrt{F[\phi,\psi]}$ defines a metric~\cite{NielsenChuang}, i.e., the fidelity is a measure for the distance of two states.
Importantly, this metric is defined in terms of \emph{states} and it is thus agnostic towards whether the Hamiltonian governing the states' evolution is Hermitian or not.

Most of the understanding of quantum geometry is limited to closed systems described by a (Hermitian) Hamiltonian.
However, the behavior of a wide range of systems~\cite{Ashida:2020fo}, e.g., open quantum systems subject to dissipation~\cite{Prosen:2008dw,Rotter:2009fr} and photonic systems with gain and loss~\cite{MartinezAlvarev:2018kg,ElGanainy:2019ie,Ozawa:2019ij}, can be described in terms of an effective non-Hermitian Hamiltonian.
Close to exceptional points, which are unique to non-Hermitian systems~\cite{Bergholtz:2021kc}, the Fubini-Study metric diverges~\cite{Solnyshkov:2021jn,Cuerda:2024ej}, as has been confirmed experimentally~\cite{Liao:2021dd,Cuerda:2024jr}.

In this work, we show how the distance of left and right eigenstates of non-Hermitian Hamiltonians gives rise to an exclusively non-Hermitian geometric term: The gauge-invariant difference between two Berry connections.
The starting point for our considerations is the fidelity between a left eigenstate and a perturbed right eigenstate, both parameterized by $\lambda$.
To lowest order in a perturbation $d \lambda$, the fidelity between these states equals
\begin{equation}
 F [ \psi_n^L (\lambda),\psi_n^R (\lambda+d\lambda) ] = \frac{ 1 - 2 d \lambda_\mu \Im \left[ \mathcal{A}^{RR}_{n,\mu} - \mathcal{A}^{LR}_{n,\mu} \right]}{\braket{\psi_n^R|\psi_n^R}\braket{\psi_n^L|\psi_n^L}}
 \label{eq:distance}
\end{equation}
with the Berry connection $\mathcal{A}^{\alpha\beta}_{n,\mu} = i \braket{\psi_n^\alpha| \partial_\mu \psi_n^\beta} /\braket{\psi_n^\alpha|\psi_n^\beta}$, and where $\ket{\psi_n^\alpha}$ with $\alpha \in L,R$ are left and right eigenstates of $H\neq H^\dagger$, satisfying $H^\dagger \ket{\psi_n^L} = E_n^* \ket{\psi_n^L}$ and $ H \ket{\psi_n^R} = E_n \ket{\psi_n^R}$, respectively.
(We give higher-order corrections in the Supplemental Material~\cite{supplemental}.)
We choose a biorthogonal normalization $\braket{\psi_n^L|\psi_{n'}^R} = \delta_{nn'}$, and denote $\braket{\psi_{n}^R|\psi_{n'}^R} = I_{nn'}$ with the positive-definite Gramian $I_{nn'}$.

The gauge-invariant~\cite{Silberstein:2020hi} difference $\mathcal{Q}_{n,\mu} = \mathcal{A}^{RR}_{n,\mu} - \mathcal{A}^{LR}_{n,\mu}$ is the central object of this study.
Previous works have highlighted the role of $\mathcal{Q}_{n,\mu}$ in semiclassical wave packet dynamics in non-Hermitian systems: Its real part gives an additional anomalous contribution to the velocity~\cite{Xu:2017bl,Silberstein:2020hi,hu2024role} of wave packets, and its imaginary part an anomalous contribution to adiabatic amplification~\cite{Silberstein:2020hi,Singhal:2023gl,Ozawa2024}.
Here, we show that $\mathcal{Q}_{n,\mu}$ contributes to non-Hermitian response functions and gives rise to terms that can be interpreted as violations of fluctuation-dissipation relations~\cite{Callen:1951be,Kubo:1957cl,Kubo:1966dq}.

The Berry connection difference $\mathcal{Q}_{n,\mu}$ also arises when considering the infinitesimal $\ket{ d \tilde{\psi}_{n,\perp}^R} = (1-P_n^{RL}) \ket{ d \psi^R_n}$, where $P_n^{\alpha\beta} = \ket{\psi_n^\alpha} \bra{\psi_n^\beta}/\braket{\psi_n^\alpha|\psi_n^\beta}$ is a projector onto the $n^\mathrm{th}$ state.
Importantly, $\ket{ d \tilde{\psi}_{n,\perp}^R}$ is orthogonal only to $\ket{\psi_n^L}$~\cite{Sun:2019en}, but not to $\ket{\psi_n^R}$; instead, their overlap equals the Berry connection difference,
\begin{equation}
 i \frac{\braket{\psi_n^R| d \tilde{\psi}^R_{n}}}{\braket{\psi_n^R|\psi_n^R}} = i \frac{\braket{\psi_n^R|d \psi_n^R}}{\braket{\psi_n^R|\psi_n^R}} - i \braket{\psi_n^L|d \psi_n^R} .
\end{equation}
From the component $\ket{ d \psi^R_{n,\perp}} = (1-P_n^{RR}) \ket{d \psi_n^R}$ orthogonal to $\ket{\psi_n^R}$~\cite{Braunstein:1994jl} we recover the standard form of the quantum geometric tensor,
\begin{equation}
 \frac{ \braket{ d \psi^R_{n,\perp} | d \psi^R_{n,\perp}} }{\braket{\psi_n^R|\psi_n^R}} = \frac{\braket{d\psi_n^R|d\psi_n^R}}{\braket{\psi_n^R|\psi_n^R}} - \frac{\braket{d\psi_n^R|\psi_n^R} \braket{\psi_n^R|d\psi_n^R}}{\braket{\psi_n^R|\psi_n^R}^2} .
\end{equation}

Previous works found different generalizations to encompass both left and right eigenstates into definitions of the geometry of quantum states~\cite{Cui:2012kg,Brody:2013et,Sun:2019en,Hu:2024kw,Chen:2024ij,orlov2024adiabatic}.
When considering infinitesimal changes of projectors $P_n^{RL} \to P_n^{RL} + dP_n^{RL}$, corrections to the fidelity between $P_n^{RL}$ and the perturbed projector are quadratic, leading to expressions analogous to the Hermitian Fubini-Study metric~\cite{Cui:2012kg,Brody:2013et,Sun:2019en}.
This generalization also arises from $\braket{ d \tilde{\psi}^L_{n} | d \tilde{\psi}^R_{n}}$ with $\ket{ d \tilde{\psi}^L_{n}} = (1-P_n^{LR}) \ket{ d \psi^L_n}$~\cite{Sun:2019en,Zhu:2021gy}.
It has been successfully applied to energy shifts in nonequilibrium dynamics~\cite{Cai:2019hc}, quantum speed limits~\cite{Sun:2019en,Hornedal2024}, and, using a multiband version, phase transitions~\cite{orlov2024adiabatic}.
In this work, however, we consider positive semi-definite density matrices $\rho$ and expectation values of the form $\tr[\dots \rho]/\tr[ \rho]$~\cite{Meden:2023ju,Bandyopadhyay2024}, which generally yields different results than biorthogonal expectation values~\cite{Brody:2014jv} (i.e., expectation values defined with respect to $P_n^{RL}$) or definitions including a time-dependent metric~\cite{Frith2019}.

We start by discussing the properties of $\mathcal{Q}_{n,\mu}$ before turning to its contribution to response functions.
We first note that for nondegenerate states
\begin{equation}
 \mathcal{Q}_{n,\mu} = i \sum_{n'\neq n} \frac{I_{nn'}}{I_{nn}} \frac{\braket{\psi_{n'}^L|\partial_\mu H |\psi_n^R}}{E_n - E_{n'}} .
 \label{eq:Q_expansion}
\end{equation}
Furthermore, for systems where states $\ket{\psi_{n}^R}$ are related to $\mathcal{K} \ket{\psi_{n}^R}$ (with complex conjugation $\mathcal{K}$) via a unitary transformation, as true for $PT$-symmetric systems~\cite{Bender:1998bw} with unbroken $PT$ symmetry~\cite{Bender:2005cd}, we find
that $\Re \mathcal{Q}_{n,\mu} =0$, resulting in a vanishing anomalous velocity in these systems~\cite{Silberstein:2020hi}.
In contrast, generally $\Im \mathcal{Q}_{n,\mu} \neq 0$.

In the subsequent discussion, we focus on translationally invariant systems governed by a non-Hermitian $\mathcal{H}_\mathbf{k}$ with complex energies $\varepsilon_{n\mathbf{k}}$.
We consider wave packets initialized in the $n$th band and defined in terms of right eigenstates $\ket{\psi_{n\mathbf{k}}^R}$,
\begin{equation}
 \ket{W_0} = \int_{\mathbf{k}} w_{\mathbf{k}} \ket{\psi_{n\mathbf{k}}^R}
 \label{eq:wave_packet}
\end{equation}
where $\int_\mathbf{k}$ denotes a normalized integral or sum over the whole Brillouin zone~\footnote{In $d$-dimensional systems, we use either $\int_\mathbf{k} = L^{-d} \sum_\mathbf{k}$ for periodic boundary conditions with linear dimension $L$, or $\int_\mathbf{k} = (2\pi)^{-d} \int d^d \mathbf{k}$ for an infinite system.}.
We choose the weights $w_\mathbf{k}$ such that the wave packet is localized in both real and momentum space with central position $\mathbf{x}_c = \langle \hat{\mathbf{x}} \rangle_0$ and momentum $\mathbf{k}_c = \langle \hat{\mathbf{k}} \rangle_0$ with the expectation value $\langle \dots \rangle_0 = \braket{W_0|\dots|W_0}/\braket{W_0|W_0}$.
We approximate $\langle f(\hat{\mathbf{k}}) \rangle_0 = f(\mathbf{k}_c)$ for generic functions $f(\mathbf{k})$, neglecting higher moments in momentum. For weights $w_\mathbf{k} = |w_\mathbf{k}| e^{i \varphi_\mathbf{k}}$, the central position
\begin{equation}
 \mathbf{x}_c = \Re [\mathcal{A}_{i}^{RR} (\mathbf{k}_c)] - \partial_{k_i} \varphi_{\mathbf{k}_c}
 \label{eq:central_position}
\end{equation}
where $\mathcal{A}_{i}^{RR} (\mathbf{k}) = i \braket{ u_{n\mathbf{k}}^R | \partial_{k_i} u_{n\mathbf{k}}^R}/\braket{u_{n\mathbf{k}}^R|u_{n\mathbf{k}}^R}$ is the momentum-space Berry connection  with the lattice-periodic part of the wave functions $\ket{u_{n\mathbf{k}}^\alpha} = e^{-i \mathbf{k} \cdot \hat{\mathbf{x}}} \ket{\psi_{n\mathbf{k}}^\alpha}$.

The Fubini-Study metric gives a lower bound on the wave packet's real-space localization~\cite{supplemental}
\begin{equation}
 \langle \hat{x}_i \hat{x}_j \rangle_c = \Re [ q_{ij} (\mathbf{k}_c) ] + \frac{1}{4} \sigma_i \sigma_j \delta_{ij}
 \label{eq:localization}
\end{equation}
with the cumulant $\langle \hat{a} \hat{b} \rangle_c := \langle \hat{a} \hat{b} \rangle_0 - \langle \hat{a} \rangle_0 \langle \hat{b} \rangle_0$ and where
\begin{equation}
 q_{ij} (\mathbf{k}) = \frac{\braket{\partial_i u_{n\mathbf{k}}^R|\partial_j u_{n\mathbf{k}}^R}}{\braket{u_{n\mathbf{k}}^R|u_{n\mathbf{k}}^R}} - \frac{\braket{\partial_i u_{n\mathbf{k}}^R|u_{n\mathbf{k}}^R} \braket{u_{n\mathbf{k}}^R|\partial_j u_{n\mathbf{k}}^R} }{\braket{u_{n\mathbf{k}}^R|u_{n\mathbf{k}}^R}\braket{u_{n\mathbf{k}}^R|u_{n\mathbf{k}}^R}}
 \label{eq:tensor}
\end{equation}
is the quantum geometric tensor.
Here $\sigma_i$ is the real-space width of the wave packet's envelope function; for Gaussian envelope functions, $|w_\mathbf{k}|^2 I_{nn} (\mathbf{k}) \propto \exp [-(1/2) \sum_i \sigma_i^2 (k_i -k_{c,i} )^2 ]$.
This result is analog to the Hermitian case, where the Fubini-Study metric provides a lower bound on the localization, important for, e.g., the construction of maximally localized Wannier functions used in electronic structure calculations~\cite{Marzari:1997co}.

In Hermitian systems, the fluctuation-dissipation theorem~\cite{Callen:1951be,Kubo:1957cl,Kubo:1966dq} links fluctuations of observables to the dissipative, i.e., imaginary, part of a corresponding response function.
The localization of ground state wave functions in insulators [Eq.~\eqref{eq:localization}] can be related to an integral over the imaginary part of the electric susceptibility~\cite{Kudinov1991,Souza:2000cj}.
Accordingly, current noise can be linked to the Fubini-Study metric~\cite{Neupert:2013eu}.
In this spirit, we now compute the electric susceptibility of generic non-Hermitian systems and show that an additional contribution arises due to the exclusively non-Hermitian contribution to the quantum geometry, i.e., the Berry curvature difference, Eq.~\eqref{eq:distance}.

The linear response of an observable $\langle \hat{a} \rangle_t = \int dt' C_{ab} (t,t') F(t')$ to a generalized force $F(t')$ that couples to a not necessarily Hermitian operator $\hat{b}(t')$ in a system governed by $H(t) \neq H^\dagger (t)$ is~\cite{Sticlet:2022en} (also cf.\ Ref.~\onlinecite{Pan:2020hp})
\begin{widetext}
\begin{align}
 C_{ab} (t,t') =& -i \Theta (t-t') \left[ \langle \hat{a} U_{t,t'} \hat{b} U_{t',t} \rangle_t - \langle \hat{a} \rangle_t \langle U_{t,t'} \hat{b} U_{t',t} \rangle_t
 - \langle U^\dagger_{t',t} \hat{b}^\dagger U^\dagger_{t,t'} \hat{a} \rangle_t + \langle U^\dagger_{t',t} \hat{b}^\dagger U^\dagger_{t,t'} \rangle_t \langle \hat{a} \rangle_t \right]
 \label{eq:response_function}
\end{align}
\end{widetext}
where $\langle \dots \rangle_t = \tr[ \dots \rho (t)]/\tr [\rho (t)]$ denotes averages with respect to the not necessarily normalized density matrix $\rho(t) = U_{t,t'} \rho (t') U_{t,t'}^\dagger$, and $U_{t,t'} = \mathcal{T} \exp [ - i \int_{t'}^t dt'' H (t'') ]$ with time ordering $\mathcal{T}$ and the unperturbed Hamiltonian $H(t)$; generally $U_{t,t'}^\dagger \neq U_{t',t}$.
We highlight two features relevant for this work: First, the unperturbed expectation value $\langle \hat{a} \rangle_t$ of $\hat{a}$ itself contributes to its response.
Second, we cannot generally switch to a Heisenberg picture~\cite{Meden:2023ju} since, due to the normalization, the time evolution is tied to \emph{states} including their gain and loss.

We now compute the electric susceptibility $\chi_{ij} (\omega)$, i.e., the response of the polarization to an external electric field~\cite{Griffiths:2017jp}, for a time-independent non-Hermiitian Hamiltonian.
Since we cannot use the Heisenberg picture and therefore cannot define a velocity operator~\cite{Meden:2023ju}, we directly compute the response $C_{ij}$ of the position $\langle \hat{x}_i \rangle_t$ to an external vector potential $A_j$ by minimal coupling $\mathcal{H}_\mathbf{k} \to \mathcal{H}_\mathbf{k} + \sum_j A_j \partial_{k_j} \mathcal{H}_\mathbf{k}$, and use $\chi_{ij} (\omega)= C_{ij} (\omega)/(i\omega)$ to obtain the susceptibility.
We resort to wave packets [Eq.~\eqref{eq:wave_packet}] to avoid problems related to ill-defined $\langle \hat{x}_i \rangle_t$ for delocalized states.
Expectation values are $\langle \dots \rangle_t = \braket{W(t)| \dots | W(t)}/\braket{W(t)|W(t)}$ with $\ket{W(t)} = e^{-i t \mathcal{H}_\mathbf{k}} \ket{W_0}$.
(Responses for fully filled bands would require extending the many-body theory of polarization~\cite{KingSmith:1993fn,Vanderbilt:1993dx} to non-Hermitian systems, which we believe depends on the context that realizes an effectively non-Hermitian time evolution. We leave this for future works.)

Starting with the general response function, Eq.~\eqref{eq:response_function} with $\hat{a} \to \hat{x}_i$ and $\hat{b} \to \partial_{k_j} \mathcal{H}_\mathbf{k}$, we find that the response function depends on the time difference $t-t'$ only~\cite{supplemental}.
(Correlation functions for generic states are not time-translation invariant~\cite{Sticlet:2022en}.)
The correlation function in frequency space is
\begin{widetext}
\begin{align}
C_{ij} (\omega)
 =& i \left( P \frac{1}{\omega} - i \pi \delta (\omega) \right) \partial_i \partial_j \Re \varepsilon_{n\mathbf{k}_c} - i \sum_{n'\neq n} \left(
 \frac{(\varepsilon_{n\mathbf{k}_c}  - \varepsilon_{n'\mathbf{k}_c}  )  f_{ij}^{nn'} (\mathbf{k}_c)   }{\omega + \varepsilon_{n\mathbf{k}_c}  -\varepsilon_{n'\mathbf{k}_c}  }
+\frac{(\varepsilon_{n\mathbf{k}_c}^*- \varepsilon_{n'\mathbf{k}_c}^*) (f_{ij}^{nn'} (\mathbf{k}_c))^*}{\omega - \varepsilon_{n\mathbf{k}_c}^*+\varepsilon_{n'\mathbf{k}_c}^*} \right) \nonumber \\
 &+ i\omega \sum_{n'\neq n} \left( \frac{ h_{ij}^{nn'} (\mathbf{k}_c)}{(\omega + \varepsilon_{n\mathbf{k}_c}-\varepsilon_{n'\mathbf{k}_c})^2} + \frac{[h_{ij}^{nn'} (\mathbf{k}_c)]^*}{(\omega - \varepsilon_{n\mathbf{k}_c}^*+\varepsilon_{n'\mathbf{k}_c}^*)^2} \right)
 \label{eq:Cij}
\end{align}
where $P$ denotes the principal value.
For convenience, we introduced two functions $f_{ij}^{nn'} (\mathbf{k})$ and $ h_{ij}^{nn'} (\mathbf{k})$ with
\begin{align}
f_{ij}^{nn'} (\mathbf{k}_c)
=& \frac{I_{nn'} (\mathbf{k}_c)}{2 I_{nn} (\mathbf{k}_c)} \left[ \braket{u_{n'\mathbf{k}_c}^L | \partial_{k_j}  u_{n \mathbf{k}_c}^R} \left( \frac{ \braket{\partial_{k_i} u_{n\mathbf{k}_c}^R |u_{n'\mathbf{k}_c}^R} - \braket{u_{n\mathbf{k}_c}^R |\partial_{k_i} u_{n'\mathbf{k}_c}^R}}{I_{nn'} (\mathbf{k}_c)} - 2 i \Re \mathcal{A}^{RR}_i (\mathbf{k}_c) \right)  - \partial_{k_i} \braket{u_{n'\mathbf{k}_c}^L |\partial_{k_j} u_{n\mathbf{k}_c}^R}  \right]
\label{eq:fij}
\end{align}
\end{widetext}
and
\begin{align}
 h_{ij}^{nn'} (\mathbf{k}_c) =& \frac{1}{2} \frac{I_{nn'} (\mathbf{k}_c)}{I_{nn} (\mathbf{k}_c)} \braket{u_{n'\mathbf{k}_c}^L | \partial_{k_j}  u_{n \mathbf{k}_c}^R} \partial_{k_i} (\varepsilon_{n\mathbf{k}_c} - \varepsilon_{n'\mathbf{k}_c} ) .
 \label{eq:hij}
\end{align}
In the following, we isolate the regular finite-frequency part $\Creg_{ij} (\omega)$ from the Drude peak with weight $\propto \partial_{k_i} \partial_{k_j} \Re \varepsilon_{n\mathbf{k}_c}$.
(Here the Drude peak has nonzero weight since wave packets correspond to partially filled bands, analogous to metallic systems~\cite{Scalapino:1993ih,Souza:2000cj}.)

Neglecting boundary terms, $C_{ij} (\omega)$ equals the optical conductivity $\sigma_{ij} (\omega)$.
In the AC limit $\omega \to 0$, we recover the anomalous velocity~\cite{Xu:2017bl,Silberstein:2020hi}
\begin{equation}
 \Creg_{ij} (\omega=0) = 2 \Im [q_{ij} ] + \partial_{k_i} \Re [ \mathcal{A}^{RR}_j - \mathcal{A}_j^{LR} ],
\end{equation}
which can also be obtained from the Kramers–Kronig relations~\cite{AltlandSimons}
\begin{equation}
 \Creg_{ij} (\omega=0) = \frac{1}{\pi} P \int_{-\infty}^\infty \frac{d \omega}{\omega} \Im   \Creg_{ij} (\omega) .
\end{equation}

In Hermitian systems, the integral of $\Im \chi_{ij} (\omega)$ over positive frequency diverges for metals and is proportional to the localization length in insulators~\cite{Kudinov1991,Scalapino:1993ih,Souza:2000cj}.
Similarly, the integral of $\Im \chi_{ij} (\omega)$ over positive frequencies diverges here, even when considering only the regularized contribution without the Drude peak.
In systems with unbroken $PT$ symmetry~\cite{Bender:1998bw}, i.e., systems with real energy eigenvalues where a symmetry operation relates eigenstates and their complex conjugates~\cite{Bender:2005cd}, we find $ \Im [f_{ij}^{nn'} (\mathbf{k}_c)]=0$.
Then, the diverging contribution vanishes, and, using that $PT$ symmetry also implies $\Im h_{ij}^{nn'} (\mathbf{k}) = 0$, we find
\begin{subequations}\begin{align}
 \int_0^\infty \frac{d\omega}{\omega} & \left. \Re  \Creg_{ij} (\omega) \right|_{PT}
 = \pi \sum_{n' \neq n} \Re f_{nn'}^{ij} (\mathbf{k}_c) \\
 &= \pi \left( \Re [q_{ij}] - \frac{1}{2} \partial_{k_i} \Im[ \mathcal{A}^{RR}_j - \mathcal{A}_j^{LR} ] \right).
\end{align}\label{eq:integral_over_ReC}\end{subequations}
This constitutes the second main result of our work:
The integral over the dissipative part of the electric susceptibility equals the fluctuations of the position operator plus the difference between left-right and right-right Berry curvature that, as we have shown, arises from the distance between left and right eigenstates.

The link between electric susceptibility and position fluctuations is not generic:
First, we needed to ignore the divergent contribution from the Drude peak and had to enforce $PT$ symmetry to make sure that $\Re \mathcal{Q}_{n,\mu} =0$.
Second, the result is only valid in the single-particle picture.
A many-body picture with filled bands, which would be necessary for a more generic link between the susceptibility and localization of states~\cite{Kudinov1991,Scalapino:1993ih,Souza:2000cj}, might result in different expressions.

\begin{figure}
\includegraphics[scale=1]{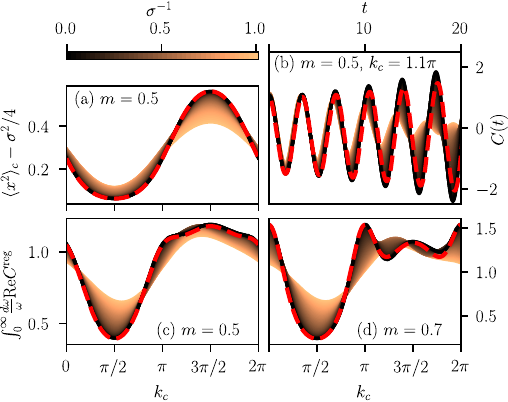}
\caption{Wave packet dynamics in a $PT$-symmetric model~\eqref{eq:hamiltonian}; colors denote different real-space spread $\sigma$. (a) Geometry-induced spread of wave packet, the red dashed line shows the real part of the geometric tensor; cf.\ Eq.~\eqref{eq:localization}. (b) Time-dependent velocity-momentum correlation for $k_c=1.1\pi$, red dashed time shows the analytical expectation for $1/\sigma \to 0$. (c) and (d) Frequency integral over $\Re C^{\mathrm{(reg)}} (\omega)/\omega$ with lifetime $\eta=0.02$ and frequency cutoff $\eta^\prime=0.2$ integrated up to a cutoff time $T=300$; the red dashed line shows the analytical expectation according to Eq.\eqref{eq:integral_over_ReC}.}
\label{fig:numerics}
\end{figure}

We demonstrate our findings by numerically computing the position-velocity response function $C (t)$ for a one-dimensional lattice toy model governed by the $PT$-symmetric Hamiltonian
\begin{equation}
 \mathcal{H}_k = \begin{pmatrix} i m \cos (k) & e^{-ik} \\ e^{ik} & -i m \cos (k) \end{pmatrix} ,
 \label{eq:hamiltonian}
\end{equation}
with real $m$; when $|m|\le 1$, all energies $\varepsilon_{k}^{(\pm)} = \pm \sqrt{1 - m^2 \cos^2 k}$ are real, and the system is in the $PT$-unbroken~\cite{Bender:2005cd} regime.
We consider a periodic system with $L$ sites, and initialize single-band Gaussian wave packets [Eq.~\eqref{eq:wave_packet}] with central momentum $k_c$ and spread $\sigma\ll L$.
In Fig.~\ref{fig:numerics}(a), we show the geometry-induced spread of wave packets initialized with different central momenta $k_c$ and envelope widths $\sigma$.
For wave packets sharply localized in momentum space, i.e., $\sigma \gg 1$, the wave packet's geometry-induced spread approaches the analytical expectation, Eq.~\eqref{eq:localization}.

The response function $C (t)$ oscillates in time around its mean value $\partial^2_k \varepsilon_{k}^{(\pm)} |_{k=k_c}$.
The magnitude of the oscillations grows as a function of time when $h(k_c)\neq 0$.
We show an example for $C (t)$ in Fig.~\ref{fig:numerics}(b), where we compare $C(t)$ with the analytical expectation for sharply peaked wave packets.
To evaluate Eq.~\eqref{eq:integral_over_ReC} numerically, we introduce a nonzero level broadening $\eta$, which in turn introduces a nonzero response $C(\omega=0)\neq 0$ at zero frequency, hence, the integral over the susceptibility $\chi_{ij} (\omega) = C(\omega)/(i\omega)$ formally diverges.
Changing $C(\omega)/\omega \to C(\omega) \omega/(\omega^2 + {\eta^\prime}^2)$ with $\eta^\prime/\eta = \mathrm{const.}$ removes this divergence.
Since we can numerically access $C(t)$ only up to a finite cutoff time $T$, and since $|C(t)|$ grows with time, we cannot numerically reach the limit $\eta\to 0$.
Instead, we set $\eta \to \eta (T)$, which introduces an exponential decay of the correlation function to counter the linear growth of $|C(t)|$ at finite integration times $T$.

We show the integrated response function for two different values of $m$ in Figs.~\ref{fig:numerics}(c) and~(d).
The integrated response function deviates from the geometry-induced  wave packet spread for nonzero $m$, with larger deviations for larger $m$.
For wave packets sharply localized in momentum space (i.e., large $\sigma$), the numerically evaluated integral matches the analytical expectation [Eq.~\eqref{eq:integral_over_ReC}].
Deviations from the analytical expectation at large $\sigma$ are due to the finite-time cutoff $T$ and the nonzero $\eta$ and $\eta^\prime$ chosen accordingly.

In this work, we showed that an exclusively non-Hermitian geometric term contributes to the electric susceptibility of $PT$-symmetric systems, which can be interpreted as a deviation from the standard fluctuation-dissipation relation~\cite{Kudinov1991,Souza:2000cj}.
The geometric term arises from the distance of left and right eigenstates and equals the gauge-independent difference between right-right and left-right Berry connection.
This term has been previously found to contribute to wave packet dynamics~\cite{Xu:2017bl,Silberstein:2020hi,hu2024role}, and adiabatic amplification of state amplitudes~\cite{Silberstein:2020hi,Singhal:2023gl,Ozawa2024}.

We explicitly demonstrate a deviation from the standard fluctuation-dissipation relation that manifests as mismatch between the integrated susceptibility and localization length.
We find this deviation for single-particle states when $PT$ symmetry is unbroken (necessary for a vanishing real part of the geometric term) and neglecting the Drude peak.

In Hermitian systems, a diverging localization length signals the transition from an insulator to a metal~\cite{Scalapino:1993ih,Souza:2000cj}.
Here, such a distinction between metals and insulators requires a many-body treatment of occupied bands, which, even for noninteracting systems, cannot directly be deduced from the single-particle behavior~\cite{Herviou:2019ku,Lee:2020ex,Alsallom:2022jz}.
Future works could thus investigate how the single-particle geometry discussed in this work affects the many-body state, e.g., mixed-state density matrices under non-Hermitian evolution (as considered in Ref.~\onlinecite{Hornedal2024}), and steady states of Lindblad operators~\cite{Prosen:2008dw,Kos:2017ew}.
This could ultimately lead to a non-Hermitian many-body generalization of the theory of polarization~\cite{Vanderbilt:1993dx,KingSmith:1993fn,Kunst:2018ku}.

While the exclusively non-Hermitian contribution to the quantum geometry that we found is a generic feature of these systems due to distinct left and right eigenstates, our discussion of the physical effects of this geometric term focused on translationally invariant systems.
We expect that the geometric term contributes to response functions beyond the specific case derived here, which can be probed using artificial gauge potentials in cold atom systems~\cite{Kolodrubetz:2017jg,Dalibard:2011gg}.

\begin{acknowledgments}
J.B. thanks Clara Wanjura for inspiring discussions.
J.B. is supported by a Leverhulme Early Career Fellowship, and the Newton Trust of the University of Cambridge.
Our simulations used resources at the Cambridge Service for Data Driven Discovery operated by the University of Cambridge Research Computing Service (\href{https://www.csd3.cam.ac.uk}{www.csd3.cam.ac.uk}), provided by Dell EMC and Intel using EPSRC Tier-2 funding via grant EP/T022159/1, and STFC DiRAC funding (\href{https://www.dirac.ac.uk}{www.dirac.ac.uk}).
R.I. is supported by the
US-Israel Binational Science Foundation (BSF) Grant
No.~2018226 and the Israel Science Foundation (ISF) grant No.~2307/24.
M.G. has been supported by the ISF and the Directorate for Defense Research and Development (DDR\&D) Grant No.~3427/21, the ISF Grant No. 113/23, and the BSF Grant No.~2020072.

\end{acknowledgments}

\appendix

\section{Higher-order corrections to distance}

Here we expand the fidelity between a left and a perturbed right eigenstate can to higher orders in the perturbation $d\lambda$.
We find that 
\begin{align}
 & F [ \psi_n^L (\lambda),\psi_n^R (\lambda+d\lambda) ] =
  \frac{1}{I_{nn} (I^{-1})_{nn}} \left[ 1 - 2 d \lambda_\mu \Im \mathcal{Q}_{n,\mu} \right.\nonumber \\
  & \left. - d \lambda_\mu d \lambda_\nu \left( \Re q^{LR}_{\mu\nu} - \partial_\nu \Im Q_{n,\mu} - 2 \Im Q_{n,\mu} \Im Q_{n,\nu} \right) \right],
\end{align}
where
\begin{align}
 q^{LR}_{\mu\nu} = \braket{\partial_\mu \psi_n^L|\partial_\nu \psi_n^R} - \braket{\partial_\mu \psi_n^L| \psi_n^R} \braket{\psi_n^L|\partial_\nu \psi_n^R}
\end{align}
is the generalization of the quantum geometric tensor found in previous works~\cite{Cui:2012kg,Brody:2013et,Sun:2019en,Hu:2024kw,Chen:2024ij,orlov2024adiabatic}.
Note that it is not only the quantum metric itself that appears in the second-order expansion, but that this term is accompanied by the left-right Berry connection difference $\mathcal{Q}_{n,\mu}$ extensively discussed in the main text.

\section{Response functions}

In the main text, we give the generic expression for the linear response function~\cite{Sticlet:2022en} of a non-Hermitian system [Eq.~\eqref{eq:response_function}].
Here, we derive in more detail the position response to an external field, i.e., the response directly linked to the electronic susceptibility.
The response function with $\hat{a} \to \hat{x}_i$ and $\hat{b} \to \hat{v}_j = \partial_{k_j} \mathcal{H}_\mathbf{k}$ can written as
\begin{equation}
  C_{ij} (t,t') = 2 \Theta (t-t') \Im C_{ij}^0 (t,t')
\end{equation}
with
\begin{align}
 C_{ij}^0 (t,t')
 =& \langle \hat{x}_i U_{t,t'} \hat{v}_j U_{t',t} \rangle_t - \langle \hat{x}_i \rangle_t \langle U_{t,t'} \hat{v}_j U_{t',t} \rangle_t .
 \label{eq:C0def}
\end{align}
Evaluating $C_{ij}^0 (t,t')$ for wave packets, i.e., $\langle \dots \rangle_t = \braket{W(t)| \dots | W(t)}/\braket{W(t)|W(t)}$, gives, using straightforward manipulations and Eq.~\eqref{eq:central_position} for position expectation values,
\begin{widetext}
\begin{align}
 C_{ij}^0 (t,t')
 = \frac{i}{2} \partial_{k_i} \partial_{k_j} \varepsilon_{n\mathbf{k}_c}  - \frac{i}{\mathcal{N}_t} \sum_{ n'\neq n} \int_\mathbf{k} & |w_\mathbf{k}|^2  e^{2 t \Im \varepsilon_{n\mathbf{k}} + i(t-t') (\varepsilon_{n \mathbf{k}}-\varepsilon_{n' \mathbf{k}})} I_{nn'} (\mathbf{k})  \braket{u_{n'\mathbf{k}}^L| \partial_{k_j}  u_{n \mathbf{k}}^R} 
 (\varepsilon_{n\mathbf{k}} - \varepsilon_{n'\mathbf{k}} )
 \label{eq:C0_simplified} \\
 & \times \left( \frac{ \braket{\partial_{k_i} u_{n\mathbf{k}}^R|u_{n'\mathbf{k}}^R}}{I_{nn'} (\mathbf{k})} + i t \partial_{k_i} \varepsilon_{n\mathbf{k}}^* + \frac{ \partial_{k_i} w_\mathbf{k}^*}{  w_\mathbf{k}^* }- i \langle \hat{x}_i \rangle_t \right) \nonumber
\end{align}
with the wave packet's time-dependent mean position $\langle \hat{x}_i \rangle_t = \Re \mathcal{A}^{RR}_i (\mathbf{k}_c) - \partial_i \varphi_{\mathbf{k}_c} + t \partial_i \Re \varepsilon_{n\mathbf{k}_c}$ and its norm $\mathcal{N}_t = \braket{W(t)|W(t)} = \mathcal{N}_0 e^{2 t \Im \varepsilon_{n\mathbf{k}_c}}$, where $\mathcal{N}_0 = \braket{W_0|W_0}$ denotes the norm of the initial wave packet at $t=0$.
We now show that the expression $C_{ij}^0 (t,t')$ depends on the time difference $t-t'$ only.
To this end, using partial integration with a generic function $q(\mathbf{k})$ [here: $q(\mathbf{k}) = (\varepsilon_{n\mathbf{k}}-\varepsilon_{n'\mathbf{k}} )  I_{nn'} (\mathbf{k}) \braket{ u_{n'\mathbf{k}}^L |\partial_{k_j} u_{n \mathbf{k}}^R}$],
\begin{align}
\int_\mathbf{k} e^{2 t \Im \varepsilon_{n\mathbf{k}} + i(t-t') (\varepsilon_{n \mathbf{k}}-\varepsilon_{n' \mathbf{k}})} q(\mathbf{k}) \left[ w_\mathbf{k} \partial_{k_i} w_\mathbf{k}^* + i t |w_\mathbf{k}|^2 \partial_{k_i} \varepsilon_{n\mathbf{k}}^* \right] 
=& \frac{i}{2} \int_\mathbf{k} e^{2 t \Im \varepsilon_{n\mathbf{k}} + i(t-t') (\varepsilon_{n \mathbf{k}}-\varepsilon_{n' \mathbf{k}})} |w_\mathbf{k}|^2 \left\{ i \partial_{k_i} q (\mathbf{k}) \right. \\
 & \left. + q(\mathbf{k}) \left[ 2 t \partial_{k_i} \Re \varepsilon_{n\mathbf{k}}  -\partial_{k_i} \varphi_\mathbf{k} - (t-t') \partial_{k_i} (\varepsilon_{n \mathbf{k}}-\varepsilon_{n' \mathbf{k}})\right] \right\} \nonumber
\end{align}
we see that the time-dependent part of $\langle \hat{x}_i \rangle_t$ in Eq.~\eqref{eq:C0_simplified} cancels the $\propto t \partial_i \Re \varepsilon_{n\mathbf{k}}$ contribution.
Performing the integration over $\mathbf{k}$ gives, using $(1/\mathcal{N}_t) \int_\mathbf{k} e^{2\Im \varepsilon_{n\mathbf{k}} t } I_{nn} (\mathbf{k}) |w_\mathbf{k}|^2 f(\mathbf{k}) = f(\mathbf{k}_c)$,
\begin{align}
 C_{ij}^0 (t,t')
 =&  \frac{i}{2} \partial_{k_i} \partial_{k_j} \varepsilon_{n\mathbf{k}_c} - i\sum_{ n'\neq n } e^{i(t-t') (\varepsilon_{n \mathbf{k}_c}-\varepsilon_{n' \mathbf{k}_c})} \left[ f_{ij}^{nn'} (\mathbf{k}_c)  (\varepsilon_{n\mathbf{k}_c} - \varepsilon_{n'\mathbf{k}_c}) - h_{ij}^{nn'} (\mathbf{k}_c) \left( 1+ i(t-t') (\varepsilon_{n\mathbf{k}_c}-\varepsilon_{n'\mathbf{k}_c})\right) \right]
\end{align}
\end{widetext}
with $f_{ij}^{nn'} (\mathbf{k})$ and $h_{ij}^{nn'} (\mathbf{k})$ defined in the main text, Eqs.~\eqref{eq:fij} and~\eqref{eq:hij}, respectively.
From $C_{ij}(t,t')= 2\Theta(t-t') \Im C_{ij}^0 (t,t')$, we obtain its Fourier transform $C_{ij}(\omega)$ that exists when $\Im [\varepsilon_{n\mathbf{k}}-\varepsilon_{n'\mathbf{k}}]>0$, i.e., when the $n$th band has the largest imaginary part (strongest gain/smallest loss), which we give in the main text, Eq.~\eqref{eq:Cij}.

We now show that
\begin{equation}
 \sum_{n\neq n'} f_{ij}^{nn'} (\mathbf{k}_c) = g_{ij} (\mathbf{k}_c) + \frac{i}{2} \partial_{k_i} \left[ \mathcal{A}_j^{RR} (\mathbf{k}_c) - \mathcal{A}_j^{LR} (\mathbf{k}_c) \right] .
\end{equation}
First note that at $\mathbf{k}_c$,
\begin{equation}
 \left. \langle \hat{x}_i \rangle_t  - t \partial_{k_i} \Re \varepsilon_{n\mathbf{k}} + \partial_{k_i} \varphi_{\mathbf{k}}  \right|_{\mathbf{k}=\mathbf{k}_c} = \Re \mathcal{A}_i^{RR} (\mathbf{k}_c) .
\end{equation}
Using the expansion Eq.~\eqref{eq:Q_expansion},
\begin{align}
& \partial_{k_i} \left( \mathcal{A}_j^{RR} - \mathcal{A}_j^{LR}\right)
= -2 \left( \mathcal{A}_j^{RR} - \mathcal{A}_j^{LR}\right) \Im \mathcal{A}_i^{RR}  \\
& + i \sum_{n'\neq n} \left[ \frac{I_{nn'}}{I_{nn}} \partial_{k_i} \braket{u_{n'\mathbf{k}}^L|\partial_{k_j} u_{n\mathbf{k}}^R} + \frac{\partial_i I_{nn'}}{I_{nn}} \braket{u_{n'\mathbf{k}}^L|\partial_{k_j} u_{n\mathbf{k}}^R} \right] \nonumber
\nonumber
\end{align}
with the identity $ (\partial_i I_{nn})/I_{nn} = 2\Im \mathcal{A}^{RR}_i$.
We thus have (with $\mathcal{Q}_j= \mathcal{A}_j^{RR} - \mathcal{A}_j^{LR}$)
\begin{align}
 & \sum_{n\neq'n} f_{ij}^{nn'} (\mathbf{k}_c)
 =\sum_{n'\neq n} \frac{\braket{\partial_i u_{n\mathbf{k}_c}^R| u_{n'\mathbf{k}_c}^R} \braket{u_{n'\mathbf{k}_c}^L | \partial_j u_{n\mathbf{k}_c}^R}}{I_{nn} (\mathbf{k}_c)} \nonumber \\
 &+\frac{i}{2} \partial_i \mathcal{Q}_j (\mathbf{k}_c)- \left[ \Re \mathcal{A}_i^{RR} (\mathbf{k}_c) -i \Im \mathcal{A}^{RR}_i (\mathbf{k}_c) \right] \mathcal{Q}_j (\mathbf{k}_c) 
\end{align}
and, since $-i \braket{\partial_i u_{n\mathbf{k}}^R|u_{n\mathbf{k}}^R}/I_{nn} = (\mathcal{A}_i^{RR})^*$,
\begin{align}
 \sum_{n\neq'n} & f_{ij}^{nn'} (\mathbf{k}_c) = \frac{i}{2} \partial_i \mathcal{Q}_j (\mathbf{k}_c) \\
 &+ \frac{\braket{\partial_i u_{n\mathbf{k}_c}^R|\partial_j u_{n\mathbf{k}_c}^R}}{\braket{u_{n\mathbf{k}_c}^R|u_{n\mathbf{k}_c}^R}}-\frac{\braket{\partial_i u_{n\mathbf{k}_c}^R|u_{n\mathbf{k}_c}^R}\braket{u_{n\mathbf{k}_c}^R| \partial_j u_{n\mathbf{k}_c}^R}}{\braket{u_{n\mathbf{k}_c}^R|u_{n\mathbf{k}_c}^R}^2} \nonumber .
\end{align}

In the main text, we show how that both real and imaginary parts of $g_{ij} + (i/2) \partial_i \mathcal{Q}_j$ arise from the response function $C_{ij}^\mathrm{(reg)} (\omega)$.
For $\omega =0$, the contribution $\propto h_{ij}^{nn'}$ vanishes and we directly have $C_{ij} (\omega=0) = 2 \sum_{nn'} \Im [f_{ij}^{nn'} (\mathbf{k}_c)]$.
The integral of the imaginary part of the electric susceptibility generally diverges, unless we demand $\Im f_{ij}^{nn'} = 0$.
As explained in the main text, this is true for $PT$-symmetric systems with unbroken $PT$ symmetry, which also implies $\Im h_{ij}^{nn'} = 0$.
Thus, the integral becomes
\begin{align}
 \int_0^\infty \frac{d\omega}{\omega} \Re C_{ij}^\mathrm{(reg)} (\omega) =& - 2 \sum_{n\neq n'}  \Re[ f_{ij}^{nn'} (\mathbf{k}_c) ] \\
 &  \times \arctan\left( \frac{\Re [\varepsilon_{n\mathbf{k}}-\varepsilon_{n'\mathbf{k}}]}{\Im [\varepsilon_{n\mathbf{k}}-\varepsilon_{n'\mathbf{k}}]} \right) \nonumber ,
\end{align}
which, for unbroken $PT$ symmetry with $\Im[\varepsilon_{n\mathbf{k}} - \varepsilon_{n'\mathbf{k'}}] \to \eta$ with an infinitesimal $\eta>0$ and assuming that the $n^\text{th}$ band is lowest in energy and separated from the rest of the spectrum, i.e., $\Re [\varepsilon_{n\mathbf{k}} - \varepsilon_{n'\mathbf{k'}}]<0$, simplifies to Eq.~\eqref{eq:integral_over_ReC} in the main text.

\section{Simulation of dynamics on a lattice}

The numerical results presented in the main text are obtained by calculating $C(t)$ for wave packets governed by a $PT$-symmetric Hamiltonian.
We compute the integral over frequencies $\omega$ directly \emph{without} Fourier transforming $C(t)$.
By introducing a level broadening $\eta$ and lower frequency cutoff $\eta^\prime$ with $\eta^\prime/\eta = \mathrm{const.}$, we replace the integral in Eq.~\eqref{eq:integral_over_ReC} by
\begin{align}
 \mathcal{I} &=  \int\limits_0^\infty \frac{d\omega}{\omega} \Re[ \Creg (\omega) ] \\
&= \lim_{\eta,\eta^\prime\to 0} \int\limits_0^\infty \frac{d\omega \omega}{\omega^2+{\eta^\prime}^2} \int\limits_0^\infty dt \Creg (t) \Re[ e^{i t (\omega+i\eta)} ]
\end{align}
where we used that $C(t)$ is real.
The integral over frequencies equals
\begin{equation}
 \int_0^\infty \frac{d\omega \omega \cos (\omega t)}{\omega^2+{\eta^\prime}^2} = -\frac{1}{2} \left(e^{\eta^\prime t} \Ei (-\eta^\prime t) + e^{-\eta^\prime t} \Ei (\eta^\prime t) \right),
\end{equation}
with the exponential integral function $\Ei (x) = -P \int_{-x}^\infty dt e^{-t}/t$.
Using partial integration, we rewrite
\begin{align}
 \mathcal{I} =& \frac{1}{2\eta^\prime} \int\limits_0^\infty dt  \left(e^{\eta^\prime t} \Ei (-\eta^\prime t) - e^{-\eta^\prime t} \Ei (\eta^\prime t) \right) \nonumber \\
 & \times \partial_t (\Creg (t)  e^{-t \eta})  \label{eq:integration}\\
 =& \frac{1}{2\eta^\prime} \int\limits_0^\infty dt e^{-t \eta} \left(e^{\eta^\prime t} \Ei (-\eta^\prime t) - e^{-\eta^\prime t} \Ei (\eta^\prime t) \right) \partial_t C (t) \nonumber .
\end{align}
In the last step, we used that the contribution to the integral from terms $\propto \partial_t e^{-\eta t}$ vanishes for all oscillating components in $\Creg (t)$. Furthermore, we replaced $\partial_t \Creg(t) = \partial_t C (t)$ since these functions differ only up to a constant.
Using Eq.~\eqref{eq:integration}, we can directly evaluate the integral without the need to subtract the constant contribution to $C(t)$ or to Fourier transform the function.

\paragraph*{Periodic boundary conditions.} To evaluate the response function [Eq.~\eqref{eq:C0def}] in a system with periodic boundary conditions, we need to ensure that the positions $x_i$ and $x_i+L$ are identified.
To this end, we first define the states $\ket{W'} = U_{tt'} v_j U_{t't} \ket{W(t)}$ and $\ket{W''} = \ket{W(t)} - i \ket{W'}$.
Using that position expectation values $\langle x_i \rangle_t := L/(2\pi) \arg [ \braket{W(t)|e^{2\pi i x_i/L}|W(t)}/\braket{W(t)|W(t)} ]$~\cite{Resta:1998hj} are well-defined for extended systems, the expression 
\begin{align}
 \Im [ & \langle x_i U_{tt'} v_j U_{t't} \rangle_t - \langle x_i \rangle_t \langle U_{tt'} v_j U_{t't} \rangle_t ] \\
  &= \langle x_i \rangle_{W''} ( \langle x_i \rangle_{W''} - \langle x_i \rangle_t ) - \langle x_i \rangle_{W'} ( \langle x_i \rangle_{W'} - \langle x_i \rangle_t ) \nonumber
\end{align}
is consistent with periodic boundary conditions, where $\langle \dots \rangle_{W'}$ and $\langle \dots \rangle_{W''}$ are averages with respect to $\ket{W'}$ and $\ket{W''}$, respectively.

\bibliography{fubini_study}

\begin{thebibliography}{83}%
\makeatletter
\providecommand \@ifxundefined [1]{%
 \@ifx{#1\undefined}
}%
\providecommand \@ifnum [1]{%
 \ifnum #1\expandafter \@firstoftwo
 \else \expandafter \@secondoftwo
 \fi
}%
\providecommand \@ifx [1]{%
 \ifx #1\expandafter \@firstoftwo
 \else \expandafter \@secondoftwo
 \fi
}%
\providecommand \natexlab [1]{#1}%
\providecommand \enquote  [1]{``#1''}%
\providecommand \bibnamefont  [1]{#1}%
\providecommand \bibfnamefont [1]{#1}%
\providecommand \citenamefont [1]{#1}%
\providecommand \href@noop [0]{\@secondoftwo}%
\providecommand \href [0]{\begingroup \@sanitize@url \@href}%
\providecommand \@href[1]{\@@startlink{#1}\@@href}%
\providecommand \@@href[1]{\endgroup#1\@@endlink}%
\providecommand \@sanitize@url [0]{\catcode `\\12\catcode `\$12\catcode
  `\&12\catcode `\#12\catcode `\^12\catcode `\_12\catcode `\%12\relax}%
\providecommand \@@startlink[1]{}%
\providecommand \@@endlink[0]{}%
\providecommand \url  [0]{\begingroup\@sanitize@url \@url }%
\providecommand \@url [1]{\endgroup\@href {#1}{\urlprefix }}%
\providecommand \urlprefix  [0]{URL }%
\providecommand \Eprint [0]{\href }%
\providecommand \doibase [0]{https://doi.org/}%
\providecommand \selectlanguage [0]{\@gobble}%
\providecommand \bibinfo  [0]{\@secondoftwo}%
\providecommand \bibfield  [0]{\@secondoftwo}%
\providecommand \translation [1]{[#1]}%
\providecommand \BibitemOpen [0]{}%
\providecommand \bibitemStop [0]{}%
\providecommand \bibitemNoStop [0]{.\EOS\space}%
\providecommand \EOS [0]{\spacefactor3000\relax}%
\providecommand \BibitemShut  [1]{\csname bibitem#1\endcsname}%
\let\auto@bib@innerbib\@empty
\bibitem [{\citenamefont {Provost}\ and\ \citenamefont
  {Vallee}(1980)}]{Provost:1980hs}%
  \BibitemOpen
  \bibfield  {author} {\bibinfo {author} {\bibfnamefont {J.~P.}\ \bibnamefont
  {Provost}}\ and\ \bibinfo {author} {\bibfnamefont {G.}~\bibnamefont
  {Vallee}},\ }\bibfield  {title} {\bibinfo {title} {{Riemannian structure on
  manifolds of quantum states}},\ }\href {https://doi.org/10.1007/BF02193559}
  {\bibfield  {journal} {\bibinfo  {journal} {Communications in Mathematical
  Physics}\ }\textbf {\bibinfo {volume} {76}},\ \bibinfo {pages} {289}
  (\bibinfo {year} {1980})}\BibitemShut {NoStop}%
\bibitem [{\citenamefont {Berry}(1989)}]{Berry:1989es}%
  \BibitemOpen
  \bibfield  {author} {\bibinfo {author} {\bibfnamefont {M.~V.}\ \bibnamefont
  {Berry}},\ }\bibfield  {title} {\bibinfo {title} {The quantum phase, five
  years after},\ }in\ \href {https://doi.org/10.1142/0613} {\emph {\bibinfo
  {booktitle} {Geometric Phases in Physics}}},\ \bibinfo {editor} {edited by\
  \bibinfo {editor} {\bibfnamefont {F.}~\bibnamefont {Wilczek}}\ and\ \bibinfo
  {editor} {\bibfnamefont {A.}~\bibnamefont {Shapere}}}\ (\bibinfo  {publisher}
  {World Scientific},\ \bibinfo {address} {Singapore},\ \bibinfo {year}
  {1989})\ pp.\ \bibinfo {pages} {7--28}\BibitemShut {NoStop}%
\bibitem [{\citenamefont {T{\"{o}}rm{\"{a}}}(2023)}]{Torma:2023jw}%
  \BibitemOpen
  \bibfield  {author} {\bibinfo {author} {\bibfnamefont {P.}~\bibnamefont
  {T{\"{o}}rm{\"{a}}}},\ }\bibfield  {title} {\bibinfo {title} {{Essay: Where
  Can Quantum Geometry Lead Us?}},\ }\href
  {https://doi.org/10.1103/PhysRevLett.131.240001} {\bibfield  {journal}
  {\bibinfo  {journal} {Physical Review Letters}\ }\textbf {\bibinfo {volume}
  {131}},\ \bibinfo {pages} {240001} (\bibinfo {year} {2023})}\BibitemShut
  {NoStop}%
\bibitem [{\citenamefont {Berry}(1984)}]{Berry:1984ka}%
  \BibitemOpen
  \bibfield  {author} {\bibinfo {author} {\bibfnamefont {M.~V.}\ \bibnamefont
  {Berry}},\ }\bibfield  {title} {\bibinfo {title} {{Quantal phase factors
  accompanying adiabatic changes}},\ }\href
  {https://doi.org/10.1098/rspa.1984.0023} {\bibfield  {journal} {\bibinfo
  {journal} {Proceedings of the Royal Society of London. A. Mathematical and
  Physical Sciences}\ }\textbf {\bibinfo {volume} {392}},\ \bibinfo {pages}
  {45} (\bibinfo {year} {1984})}\BibitemShut {NoStop}%
\bibitem [{\citenamefont {Xiao}\ \emph {et~al.}(2010)\citenamefont {Xiao},
  \citenamefont {Chang},\ and\ \citenamefont {Niu}}]{Xiao:2010kw}%
  \BibitemOpen
  \bibfield  {author} {\bibinfo {author} {\bibfnamefont {D.}~\bibnamefont
  {Xiao}}, \bibinfo {author} {\bibfnamefont {M.-C.}\ \bibnamefont {Chang}},\
  and\ \bibinfo {author} {\bibfnamefont {Q.}~\bibnamefont {Niu}},\ }\bibfield
  {title} {\bibinfo {title} {{Berry phase effects on electronic properties}},\
  }\href {https://doi.org/10.1103/RevModPhys.82.1959} {\bibfield  {journal}
  {\bibinfo  {journal} {Reviews of Modern Physics}\ }\textbf {\bibinfo {volume}
  {82}},\ \bibinfo {pages} {1959} (\bibinfo {year} {2010})}\BibitemShut
  {NoStop}%
\bibitem [{\citenamefont {Heslot}(1985)}]{Heslot:1985bw}%
  \BibitemOpen
  \bibfield  {author} {\bibinfo {author} {\bibfnamefont {A.}~\bibnamefont
  {Heslot}},\ }\bibfield  {title} {\bibinfo {title} {{Quantum mechanics as a
  classical theory}},\ }\href {https://doi.org/10.1103/PhysRevD.31.1341}
  {\bibfield  {journal} {\bibinfo  {journal} {Physical Review D}\ }\textbf
  {\bibinfo {volume} {31}},\ \bibinfo {pages} {1341} (\bibinfo {year}
  {1985})}\BibitemShut {NoStop}%
\bibitem [{\citenamefont {Ashtekar}\ and\ \citenamefont
  {Schilling}(1999)}]{Ashtekar:1999iy}%
  \BibitemOpen
  \bibfield  {author} {\bibinfo {author} {\bibfnamefont {A.}~\bibnamefont
  {Ashtekar}}\ and\ \bibinfo {author} {\bibfnamefont {T.~A.}\ \bibnamefont
  {Schilling}},\ }\bibfield  {title} {\bibinfo {title} {{Geometrical
  Formulation of Quantum Mechanics}},\ }in\ \href
  {https://doi.org/10.1007/978-1-4612-1422-9_3} {\emph {\bibinfo {booktitle}
  {On Einstein's Path}}},\ \bibinfo {editor} {edited by\ \bibinfo {editor}
  {\bibfnamefont {A.}~\bibnamefont {Harvey}}}\ (\bibinfo  {publisher}
  {Springer},\ \bibinfo {address} {New York, NY},\ \bibinfo {year} {1999})\
  pp.\ \bibinfo {pages} {23--65}\BibitemShut {NoStop}%
\bibitem [{\citenamefont {Facchi}\ \emph {et~al.}(2010)\citenamefont {Facchi},
  \citenamefont {Kulkarni}, \citenamefont {Man'ko}, \citenamefont {Marmo},
  \citenamefont {Sudarshan},\ and\ \citenamefont {Ventriglia}}]{Facchi:2010hl}%
  \BibitemOpen
  \bibfield  {author} {\bibinfo {author} {\bibfnamefont {P.}~\bibnamefont
  {Facchi}}, \bibinfo {author} {\bibfnamefont {R.}~\bibnamefont {Kulkarni}},
  \bibinfo {author} {\bibfnamefont {V.}~\bibnamefont {Man'ko}}, \bibinfo
  {author} {\bibfnamefont {G.}~\bibnamefont {Marmo}}, \bibinfo {author}
  {\bibfnamefont {E.}~\bibnamefont {Sudarshan}},\ and\ \bibinfo {author}
  {\bibfnamefont {F.}~\bibnamefont {Ventriglia}},\ }\bibfield  {title}
  {\bibinfo {title} {{Classical and quantum Fisher information in the
  geometrical formulation of quantum mechanics}},\ }\href
  {https://doi.org/10.1016/j.physleta.2010.10.005} {\bibfield  {journal}
  {\bibinfo  {journal} {Physics Letters A}\ }\textbf {\bibinfo {volume}
  {374}},\ \bibinfo {pages} {4801} (\bibinfo {year} {2010})}\BibitemShut
  {NoStop}%
\bibitem [{\citenamefont {Chang}\ and\ \citenamefont
  {Niu}(1995)}]{Chang:1995gk}%
  \BibitemOpen
  \bibfield  {author} {\bibinfo {author} {\bibfnamefont {M.-C.}\ \bibnamefont
  {Chang}}\ and\ \bibinfo {author} {\bibfnamefont {Q.}~\bibnamefont {Niu}},\
  }\bibfield  {title} {\bibinfo {title} {{Berry Phase, Hyperorbits, and the
  Hofstadter Spectrum}},\ }\href {https://doi.org/10.1103/PhysRevLett.75.1348}
  {\bibfield  {journal} {\bibinfo  {journal} {Physical Review Letters}\
  }\textbf {\bibinfo {volume} {75}},\ \bibinfo {pages} {1348} (\bibinfo {year}
  {1995})}\BibitemShut {NoStop}%
\bibitem [{\citenamefont {Chang}\ and\ \citenamefont
  {Niu}(1996)}]{Chang:1996gv}%
  \BibitemOpen
  \bibfield  {author} {\bibinfo {author} {\bibfnamefont {M.-C.}\ \bibnamefont
  {Chang}}\ and\ \bibinfo {author} {\bibfnamefont {Q.}~\bibnamefont {Niu}},\
  }\bibfield  {title} {\bibinfo {title} {{Berry phase, hyperorbits, and the
  Hofstadter spectrum: Semiclassical dynamics in magnetic Bloch bands}},\
  }\href {https://doi.org/10.1103/PhysRevB.53.7010} {\bibfield  {journal}
  {\bibinfo  {journal} {Physical Review B}\ }\textbf {\bibinfo {volume} {53}},\
  \bibinfo {pages} {7010} (\bibinfo {year} {1996})}\BibitemShut {NoStop}%
\bibitem [{\citenamefont {Thouless}\ \emph {et~al.}(1982)\citenamefont
  {Thouless}, \citenamefont {Kohmoto}, \citenamefont {Nightingale},\ and\
  \citenamefont {den Nijs}}]{Thouless:1982kq}%
  \BibitemOpen
  \bibfield  {author} {\bibinfo {author} {\bibfnamefont {D.~J.}\ \bibnamefont
  {Thouless}}, \bibinfo {author} {\bibfnamefont {M.}~\bibnamefont {Kohmoto}},
  \bibinfo {author} {\bibfnamefont {M.~P.}\ \bibnamefont {Nightingale}},\ and\
  \bibinfo {author} {\bibfnamefont {M.}~\bibnamefont {den Nijs}},\ }\bibfield
  {title} {\bibinfo {title} {{Quantized Hall Conductance in a Two-Dimensional
  Periodic Potential}},\ }\href {https://doi.org/10.1103/PhysRevLett.49.405}
  {\bibfield  {journal} {\bibinfo  {journal} {Physical Review Letters}\
  }\textbf {\bibinfo {volume} {49}},\ \bibinfo {pages} {405} (\bibinfo {year}
  {1982})}\BibitemShut {NoStop}%
\bibitem [{\citenamefont {Thouless}(1983)}]{Thouless:1983hb}%
  \BibitemOpen
  \bibfield  {author} {\bibinfo {author} {\bibfnamefont {D.~J.}\ \bibnamefont
  {Thouless}},\ }\bibfield  {title} {\bibinfo {title} {{Quantization of
  particle transport}},\ }\href {https://doi.org/10.1103/PhysRevB.27.6083}
  {\bibfield  {journal} {\bibinfo  {journal} {Physical Review B}\ }\textbf
  {\bibinfo {volume} {27}},\ \bibinfo {pages} {6083} (\bibinfo {year}
  {1983})}\BibitemShut {NoStop}%
\bibitem [{\citenamefont {Study}(1905)}]{Study:1905ku}%
  \BibitemOpen
  \bibfield  {author} {\bibinfo {author} {\bibfnamefont {E.}~\bibnamefont
  {Study}},\ }\bibfield  {title} {\bibinfo {title} {{K{\"{u}}rzeste Wege im
  komplexen Gebiet}},\ }\href {https://doi.org/10.1007/BF01457616} {\bibfield
  {journal} {\bibinfo  {journal} {Mathematische Annalen}\ }\textbf {\bibinfo
  {volume} {60}},\ \bibinfo {pages} {321} (\bibinfo {year} {1905})}\BibitemShut
  {NoStop}%
\bibitem [{Note1()}]{Note1}%
  \BibitemOpen
  \bibinfo {note} {Naming conventions of these terms vary~\cite
  {Kolodrubetz:2017jg}. Throughout this work, we refer to the complex object
  $q_{\mu \nu }$ as the quantum geometric tensor, its real part the
  Fubini-Study metric~\cite {Study:1905ku,Provost:1980hs,Anandan:1990dx} (often
  called the quantum metric~\cite {Neupert:2013eu,Kolodrubetz:2013bf}), and
  twice its imaginary part the Berry curvature.}\BibitemShut {Stop}%
\bibitem [{\citenamefont {Anandan}\ and\ \citenamefont
  {Aharonov}(1990)}]{Anandan:1990dx}%
  \BibitemOpen
  \bibfield  {author} {\bibinfo {author} {\bibfnamefont {J.}~\bibnamefont
  {Anandan}}\ and\ \bibinfo {author} {\bibfnamefont {Y.}~\bibnamefont
  {Aharonov}},\ }\bibfield  {title} {\bibinfo {title} {{Geometry of quantum
  evolution}},\ }\href {https://doi.org/10.1103/PhysRevLett.65.1697} {\bibfield
   {journal} {\bibinfo  {journal} {Physical Review Letters}\ }\textbf {\bibinfo
  {volume} {65}},\ \bibinfo {pages} {1697} (\bibinfo {year}
  {1990})}\BibitemShut {NoStop}%
\bibitem [{\citenamefont {{De Grandi}}\ \emph {et~al.}(2011)\citenamefont {{De
  Grandi}}, \citenamefont {Polkovnikov},\ and\ \citenamefont
  {Sandvik}}]{DeGrandi:2011kq}%
  \BibitemOpen
  \bibfield  {author} {\bibinfo {author} {\bibfnamefont {C.}~\bibnamefont {{De
  Grandi}}}, \bibinfo {author} {\bibfnamefont {A.}~\bibnamefont
  {Polkovnikov}},\ and\ \bibinfo {author} {\bibfnamefont {A.~W.}\ \bibnamefont
  {Sandvik}},\ }\bibfield  {title} {\bibinfo {title} {{Universal nonequilibrium
  quantum dynamics in imaginary time}},\ }\href
  {https://doi.org/10.1103/PhysRevB.84.224303} {\bibfield  {journal} {\bibinfo
  {journal} {Physical Review B}\ }\textbf {\bibinfo {volume} {84}},\ \bibinfo
  {pages} {224303} (\bibinfo {year} {2011})}\BibitemShut {NoStop}%
\bibitem [{\citenamefont {Yang}\ \emph {et~al.}(2008)\citenamefont {Yang},
  \citenamefont {Gu}, \citenamefont {Sun},\ and\ \citenamefont
  {Lin}}]{Yang:2008cl}%
  \BibitemOpen
  \bibfield  {author} {\bibinfo {author} {\bibfnamefont {S.}~\bibnamefont
  {Yang}}, \bibinfo {author} {\bibfnamefont {S.-J.}\ \bibnamefont {Gu}},
  \bibinfo {author} {\bibfnamefont {C.-P.}\ \bibnamefont {Sun}},\ and\ \bibinfo
  {author} {\bibfnamefont {H.-Q.}\ \bibnamefont {Lin}},\ }\bibfield  {title}
  {\bibinfo {title} {{Fidelity susceptibility and long-range correlation in the
  Kitaev honeycomb model}},\ }\href
  {https://doi.org/10.1103/PhysRevA.78.012304} {\bibfield  {journal} {\bibinfo
  {journal} {Physical Review A}\ }\textbf {\bibinfo {volume} {78}},\ \bibinfo
  {pages} {012304} (\bibinfo {year} {2008})}\BibitemShut {NoStop}%
\bibitem [{\citenamefont {Garnerone}\ \emph {et~al.}(2009)\citenamefont
  {Garnerone}, \citenamefont {Abasto}, \citenamefont {Haas},\ and\
  \citenamefont {Zanardi}}]{Garnerone:2009ho}%
  \BibitemOpen
  \bibfield  {author} {\bibinfo {author} {\bibfnamefont {S.}~\bibnamefont
  {Garnerone}}, \bibinfo {author} {\bibfnamefont {D.}~\bibnamefont {Abasto}},
  \bibinfo {author} {\bibfnamefont {S.}~\bibnamefont {Haas}},\ and\ \bibinfo
  {author} {\bibfnamefont {P.}~\bibnamefont {Zanardi}},\ }\bibfield  {title}
  {\bibinfo {title} {{Fidelity in topological quantum phases of matter}},\
  }\href {https://doi.org/10.1103/PhysRevA.79.032302} {\bibfield  {journal}
  {\bibinfo  {journal} {Physical Review A}\ }\textbf {\bibinfo {volume} {79}},\
  \bibinfo {pages} {032302} (\bibinfo {year} {2009})}\BibitemShut {NoStop}%
\bibitem [{\citenamefont {Kudinov}(1991)}]{Kudinov1991}%
  \BibitemOpen
  \bibfield  {author} {\bibinfo {author} {\bibfnamefont {E.~K.}\ \bibnamefont
  {Kudinov}},\ }\bibfield  {title} {\bibinfo {title} {{Difference between
  insulating and conducting states}},\ }\href@noop {} {\bibfield  {journal}
  {\bibinfo  {journal} {Fis. Tverd. Tela}\ }\textbf {\bibinfo {volume} {33}},\
  \bibinfo {pages} {2306} (\bibinfo {year} {1991})},\ \bibinfo {note} {[Sov.\
  Phys.\ Solid State 33, 1299 (1991)]}\BibitemShut {NoStop}%
\bibitem [{\citenamefont {Marzari}\ and\ \citenamefont
  {Vanderbilt}(1997)}]{Marzari:1997co}%
  \BibitemOpen
  \bibfield  {author} {\bibinfo {author} {\bibfnamefont {N.}~\bibnamefont
  {Marzari}}\ and\ \bibinfo {author} {\bibfnamefont {D.}~\bibnamefont
  {Vanderbilt}},\ }\bibfield  {title} {\bibinfo {title} {{Maximally localized
  generalized Wannier functions for composite energy bands}},\ }\href
  {https://doi.org/10.1103/PhysRevB.56.12847} {\bibfield  {journal} {\bibinfo
  {journal} {Physical Review B}\ }\textbf {\bibinfo {volume} {56}},\ \bibinfo
  {pages} {12847} (\bibinfo {year} {1997})}\BibitemShut {NoStop}%
\bibitem [{\citenamefont {Resta}\ and\ \citenamefont
  {Sorella}(1999)}]{Resta:1999kl}%
  \BibitemOpen
  \bibfield  {author} {\bibinfo {author} {\bibfnamefont {R.}~\bibnamefont
  {Resta}}\ and\ \bibinfo {author} {\bibfnamefont {S.}~\bibnamefont
  {Sorella}},\ }\bibfield  {title} {\bibinfo {title} {{Electron Localization in
  the Insulating State}},\ }\href {https://doi.org/10.1103/PhysRevLett.82.370}
  {\bibfield  {journal} {\bibinfo  {journal} {Physical Review Letters}\
  }\textbf {\bibinfo {volume} {82}},\ \bibinfo {pages} {370} (\bibinfo {year}
  {1999})}\BibitemShut {NoStop}%
\bibitem [{\citenamefont {Neupert}\ \emph {et~al.}(2013)\citenamefont
  {Neupert}, \citenamefont {Chamon},\ and\ \citenamefont
  {Mudry}}]{Neupert:2013eu}%
  \BibitemOpen
  \bibfield  {author} {\bibinfo {author} {\bibfnamefont {T.}~\bibnamefont
  {Neupert}}, \bibinfo {author} {\bibfnamefont {C.}~\bibnamefont {Chamon}},\
  and\ \bibinfo {author} {\bibfnamefont {C.}~\bibnamefont {Mudry}},\ }\bibfield
   {title} {\bibinfo {title} {{Measuring the quantum geometry of Bloch bands
  with current noise}},\ }\href {https://doi.org/10.1103/PhysRevB.87.245103}
  {\bibfield  {journal} {\bibinfo  {journal} {Phys. Rev. B}\ }\textbf {\bibinfo
  {volume} {87}},\ \bibinfo {pages} {245103} (\bibinfo {year}
  {2013})}\BibitemShut {NoStop}%
\bibitem [{\citenamefont {Kivelson}(1982)}]{Kivelson:1982fr}%
  \BibitemOpen
  \bibfield  {author} {\bibinfo {author} {\bibfnamefont {S.}~\bibnamefont
  {Kivelson}},\ }\bibfield  {title} {\bibinfo {title} {{Wannier functions in
  one-dimensional disordered systems: Application to fractionally charged
  solitons}},\ }\href {https://doi.org/10.1103/PhysRevB.26.4269} {\bibfield
  {journal} {\bibinfo  {journal} {Physical Review B}\ }\textbf {\bibinfo
  {volume} {26}},\ \bibinfo {pages} {4269} (\bibinfo {year}
  {1982})}\BibitemShut {NoStop}%
\bibitem [{\citenamefont {Onishi}\ and\ \citenamefont
  {Fu}(2024)}]{Onishi:2024gl}%
  \BibitemOpen
  \bibfield  {author} {\bibinfo {author} {\bibfnamefont {Y.}~\bibnamefont
  {Onishi}}\ and\ \bibinfo {author} {\bibfnamefont {L.}~\bibnamefont {Fu}},\
  }\bibfield  {title} {\bibinfo {title} {{Fundamental Bound on Topological
  Gap}},\ }\href {https://doi.org/10.1103/PhysRevX.14.011052} {\bibfield
  {journal} {\bibinfo  {journal} {Physical Review X}\ }\textbf {\bibinfo
  {volume} {14}},\ \bibinfo {pages} {011052} (\bibinfo {year}
  {2024})}\BibitemShut {NoStop}%
\bibitem [{\citenamefont {Onishi}\ and\ \citenamefont {Fu}()}]{Onishi2024}%
  \BibitemOpen
  \bibfield  {author} {\bibinfo {author} {\bibfnamefont {Y.}~\bibnamefont
  {Onishi}}\ and\ \bibinfo {author} {\bibfnamefont {L.}~\bibnamefont {Fu}},\
  }\bibfield  {title} {\bibinfo {title} {{Quantum weight}},\ }\Eprint
  {https://arxiv.org/abs/2401.13847} {arXiv:2401.13847} \BibitemShut {NoStop}%
\bibitem [{\citenamefont {Peotta}\ and\ \citenamefont
  {T{\"{o}}rm{\"{a}}}(2015)}]{Peotta:2015gb}%
  \BibitemOpen
  \bibfield  {author} {\bibinfo {author} {\bibfnamefont {S.}~\bibnamefont
  {Peotta}}\ and\ \bibinfo {author} {\bibfnamefont {P.}~\bibnamefont
  {T{\"{o}}rm{\"{a}}}},\ }\bibfield  {title} {\bibinfo {title} {{Superfluidity
  in topologically nontrivial flat bands}},\ }\href
  {https://doi.org/10.1038/ncomms9944} {\bibfield  {journal} {\bibinfo
  {journal} {Nature Communications}\ }\textbf {\bibinfo {volume} {6}},\
  \bibinfo {pages} {8944} (\bibinfo {year} {2015})}\BibitemShut {NoStop}%
\bibitem [{\citenamefont {Gao}\ \emph {et~al.}(2014)\citenamefont {Gao},
  \citenamefont {Yang},\ and\ \citenamefont {Niu}}]{Gao:2014cn}%
  \BibitemOpen
  \bibfield  {author} {\bibinfo {author} {\bibfnamefont {Y.}~\bibnamefont
  {Gao}}, \bibinfo {author} {\bibfnamefont {S.~A.}\ \bibnamefont {Yang}},\ and\
  \bibinfo {author} {\bibfnamefont {Q.}~\bibnamefont {Niu}},\ }\bibfield
  {title} {\bibinfo {title} {{Field Induced Positional Shift of Bloch Electrons
  and Its Dynamical Implications}},\ }\href
  {https://doi.org/10.1103/PhysRevLett.112.166601} {\bibfield  {journal}
  {\bibinfo  {journal} {Physical Review Letters}\ }\textbf {\bibinfo {volume}
  {112}},\ \bibinfo {pages} {166601} (\bibinfo {year} {2014})}\BibitemShut
  {NoStop}%
\bibitem [{\citenamefont {Jozsa}(1994)}]{Jozsa:1994ik}%
  \BibitemOpen
  \bibfield  {author} {\bibinfo {author} {\bibfnamefont {R.}~\bibnamefont
  {Jozsa}},\ }\bibfield  {title} {\bibinfo {title} {{Fidelity for Mixed Quantum
  States}},\ }\href {https://doi.org/10.1080/09500349414552171} {\bibfield
  {journal} {\bibinfo  {journal} {Journal of Modern Optics}\ }\textbf {\bibinfo
  {volume} {41}},\ \bibinfo {pages} {2315} (\bibinfo {year}
  {1994})}\BibitemShut {NoStop}%
\bibitem [{\citenamefont {Nielsen}\ and\ \citenamefont
  {Chuang}(2010)}]{NielsenChuang}%
  \BibitemOpen
  \bibfield  {author} {\bibinfo {author} {\bibfnamefont {M.~A.}\ \bibnamefont
  {Nielsen}}\ and\ \bibinfo {author} {\bibfnamefont {I.~L.}\ \bibnamefont
  {Chuang}},\ }\href {https://doi.org/10.1017/CBO9780511976667} {\emph
  {\bibinfo {title} {{Quantum Computation and Quantum Information}}}}\
  (\bibinfo  {publisher} {Cambridge University Press},\ \bibinfo {address}
  {Cambridge, U.K.},\ \bibinfo {year} {2010})\BibitemShut {NoStop}%
\bibitem [{\citenamefont {Ashida}\ \emph {et~al.}(2020)\citenamefont {Ashida},
  \citenamefont {Gong},\ and\ \citenamefont {Ueda}}]{Ashida:2020fo}%
  \BibitemOpen
  \bibfield  {author} {\bibinfo {author} {\bibfnamefont {Y.}~\bibnamefont
  {Ashida}}, \bibinfo {author} {\bibfnamefont {Z.}~\bibnamefont {Gong}},\ and\
  \bibinfo {author} {\bibfnamefont {M.}~\bibnamefont {Ueda}},\ }\bibfield
  {title} {\bibinfo {title} {{Non-Hermitian physics}},\ }\href
  {https://doi.org/10.1080/00018732.2021.1876991} {\bibfield  {journal}
  {\bibinfo  {journal} {Advances in Physics}\ }\textbf {\bibinfo {volume}
  {69}},\ \bibinfo {pages} {249} (\bibinfo {year} {2020})}\BibitemShut
  {NoStop}%
\bibitem [{\citenamefont {Prosen}(2008)}]{Prosen:2008dw}%
  \BibitemOpen
  \bibfield  {author} {\bibinfo {author} {\bibfnamefont {T.}~\bibnamefont
  {Prosen}},\ }\bibfield  {title} {\bibinfo {title} {{Third quantization: a
  general method to solve master equations for quadratic open Fermi systems}},\
  }\href {https://doi.org/10.1088/1367-2630/10/4/043026} {\bibfield  {journal}
  {\bibinfo  {journal} {New Journal of Physics}\ }\textbf {\bibinfo {volume}
  {10}},\ \bibinfo {pages} {043026} (\bibinfo {year} {2008})}\BibitemShut
  {NoStop}%
\bibitem [{\citenamefont {Rotter}(2009)}]{Rotter:2009fr}%
  \BibitemOpen
  \bibfield  {author} {\bibinfo {author} {\bibfnamefont {I.}~\bibnamefont
  {Rotter}},\ }\bibfield  {title} {\bibinfo {title} {{A non-Hermitian Hamilton
  operator and the physics of open quantum systems}},\ }\href
  {https://doi.org/10.1088/1751-8113/42/15/153001} {\bibfield  {journal}
  {\bibinfo  {journal} {Journal of Physics A: Mathematical and Theoretical}\
  }\textbf {\bibinfo {volume} {42}},\ \bibinfo {pages} {153001} (\bibinfo
  {year} {2009})}\BibitemShut {NoStop}%
\bibitem [{\citenamefont {{Martinez Alvarez}}\ \emph
  {et~al.}(2018)\citenamefont {{Martinez Alvarez}}, \citenamefont {{Barrios
  Vargas}}, \citenamefont {Berdakin},\ and\ \citenamefont {{Foa
  Torres}}}]{MartinezAlvarev:2018kg}%
  \BibitemOpen
  \bibfield  {author} {\bibinfo {author} {\bibfnamefont {V.~M.}\ \bibnamefont
  {{Martinez Alvarez}}}, \bibinfo {author} {\bibfnamefont {J.~E.}\ \bibnamefont
  {{Barrios Vargas}}}, \bibinfo {author} {\bibfnamefont {M.}~\bibnamefont
  {Berdakin}},\ and\ \bibinfo {author} {\bibfnamefont {L.~E.~F.}\ \bibnamefont
  {{Foa Torres}}},\ }\bibfield  {title} {\bibinfo {title} {{Topological states
  of non-Hermitian systems}},\ }\href
  {https://doi.org/10.1140/epjst/e2018-800091-5} {\bibfield  {journal}
  {\bibinfo  {journal} {The European Physical Journal Special Topics}\ }\textbf
  {\bibinfo {volume} {227}},\ \bibinfo {pages} {1295} (\bibinfo {year}
  {2018})}\BibitemShut {NoStop}%
\bibitem [{\citenamefont {El-Ganainy}\ \emph {et~al.}(2018)\citenamefont
  {El-Ganainy}, \citenamefont {Makris}, \citenamefont {Khajavikhan},
  \citenamefont {Musslimani}, \citenamefont {Rotter},\ and\ \citenamefont
  {Christodoulides}}]{ElGanainy:2019ie}%
  \BibitemOpen
  \bibfield  {author} {\bibinfo {author} {\bibfnamefont {R.}~\bibnamefont
  {El-Ganainy}}, \bibinfo {author} {\bibfnamefont {K.~G.}\ \bibnamefont
  {Makris}}, \bibinfo {author} {\bibfnamefont {M.}~\bibnamefont {Khajavikhan}},
  \bibinfo {author} {\bibfnamefont {Z.~H.}\ \bibnamefont {Musslimani}},
  \bibinfo {author} {\bibfnamefont {S.}~\bibnamefont {Rotter}},\ and\ \bibinfo
  {author} {\bibfnamefont {D.~N.}\ \bibnamefont {Christodoulides}},\ }\bibfield
   {title} {\bibinfo {title} {{Non-Hermitian physics and PT symmetry}},\ }\href
  {https://doi.org/10.1038/nphys4323} {\bibfield  {journal} {\bibinfo
  {journal} {Nature Physics}\ }\textbf {\bibinfo {volume} {14}},\ \bibinfo
  {pages} {11} (\bibinfo {year} {2018})}\BibitemShut {NoStop}%
\bibitem [{\citenamefont {Ozawa}\ \emph {et~al.}(2019)\citenamefont {Ozawa},
  \citenamefont {Price}, \citenamefont {Amo}, \citenamefont {Goldman},
  \citenamefont {Hafezi}, \citenamefont {Lu}, \citenamefont {Rechtsman},
  \citenamefont {Schuster}, \citenamefont {Simon}, \citenamefont {Zilberberg},\
  and\ \citenamefont {Carusotto}}]{Ozawa:2019ij}%
  \BibitemOpen
  \bibfield  {author} {\bibinfo {author} {\bibfnamefont {T.}~\bibnamefont
  {Ozawa}}, \bibinfo {author} {\bibfnamefont {H.~M.}\ \bibnamefont {Price}},
  \bibinfo {author} {\bibfnamefont {A.}~\bibnamefont {Amo}}, \bibinfo {author}
  {\bibfnamefont {N.}~\bibnamefont {Goldman}}, \bibinfo {author} {\bibfnamefont
  {M.}~\bibnamefont {Hafezi}}, \bibinfo {author} {\bibfnamefont
  {L.}~\bibnamefont {Lu}}, \bibinfo {author} {\bibfnamefont {M.~C.}\
  \bibnamefont {Rechtsman}}, \bibinfo {author} {\bibfnamefont {D.}~\bibnamefont
  {Schuster}}, \bibinfo {author} {\bibfnamefont {J.}~\bibnamefont {Simon}},
  \bibinfo {author} {\bibfnamefont {O.}~\bibnamefont {Zilberberg}},\ and\
  \bibinfo {author} {\bibfnamefont {I.}~\bibnamefont {Carusotto}},\ }\bibfield
  {title} {\bibinfo {title} {{Topological photonics}},\ }\href
  {https://doi.org/10.1103/RevModPhys.91.015006} {\bibfield  {journal}
  {\bibinfo  {journal} {Reviews of Modern Physics}\ }\textbf {\bibinfo {volume}
  {91}},\ \bibinfo {pages} {015006} (\bibinfo {year} {2019})}\BibitemShut
  {NoStop}%
\bibitem [{\citenamefont {Bergholtz}\ \emph {et~al.}(2021)\citenamefont
  {Bergholtz}, \citenamefont {Budich},\ and\ \citenamefont
  {Kunst}}]{Bergholtz:2021kc}%
  \BibitemOpen
  \bibfield  {author} {\bibinfo {author} {\bibfnamefont {E.~J.}\ \bibnamefont
  {Bergholtz}}, \bibinfo {author} {\bibfnamefont {J.~C.}\ \bibnamefont
  {Budich}},\ and\ \bibinfo {author} {\bibfnamefont {F.~K.}\ \bibnamefont
  {Kunst}},\ }\bibfield  {title} {\bibinfo {title} {{Exceptional topology of
  non-Hermitian systems}},\ }\href
  {https://doi.org/10.1103/RevModPhys.93.015005} {\bibfield  {journal}
  {\bibinfo  {journal} {Reviews of Modern Physics}\ }\textbf {\bibinfo {volume}
  {93}},\ \bibinfo {pages} {015005} (\bibinfo {year} {2021})}\BibitemShut
  {NoStop}%
\bibitem [{\citenamefont {Solnyshkov}\ \emph {et~al.}(2021)\citenamefont
  {Solnyshkov}, \citenamefont {Leblanc}, \citenamefont {Bessonart},
  \citenamefont {Nalitov}, \citenamefont {Ren}, \citenamefont {Liao},
  \citenamefont {Li},\ and\ \citenamefont {Malpuech}}]{Solnyshkov:2021jn}%
  \BibitemOpen
  \bibfield  {author} {\bibinfo {author} {\bibfnamefont {D.~D.}\ \bibnamefont
  {Solnyshkov}}, \bibinfo {author} {\bibfnamefont {C.}~\bibnamefont {Leblanc}},
  \bibinfo {author} {\bibfnamefont {L.}~\bibnamefont {Bessonart}}, \bibinfo
  {author} {\bibfnamefont {A.}~\bibnamefont {Nalitov}}, \bibinfo {author}
  {\bibfnamefont {J.}~\bibnamefont {Ren}}, \bibinfo {author} {\bibfnamefont
  {Q.}~\bibnamefont {Liao}}, \bibinfo {author} {\bibfnamefont {F.}~\bibnamefont
  {Li}},\ and\ \bibinfo {author} {\bibfnamefont {G.}~\bibnamefont {Malpuech}},\
  }\bibfield  {title} {\bibinfo {title} {{Quantum metric and wave packets at
  exceptional points in non-Hermitian systems}},\ }\href
  {https://doi.org/10.1103/PhysRevB.103.125302} {\bibfield  {journal} {\bibinfo
   {journal} {Physical Review B}\ }\textbf {\bibinfo {volume} {103}},\ \bibinfo
  {pages} {125302} (\bibinfo {year} {2021})}\BibitemShut {NoStop}%
\bibitem [{\citenamefont {Cuerda}\ \emph
  {et~al.}(2024{\natexlab{a}})\citenamefont {Cuerda}, \citenamefont {Taskinen},
  \citenamefont {K{\"{a}}llman}, \citenamefont {Grabitz},\ and\ \citenamefont
  {T{\"{o}}rm{\"{a}}}}]{Cuerda:2024ej}%
  \BibitemOpen
  \bibfield  {author} {\bibinfo {author} {\bibfnamefont {J.}~\bibnamefont
  {Cuerda}}, \bibinfo {author} {\bibfnamefont {J.~M.}\ \bibnamefont
  {Taskinen}}, \bibinfo {author} {\bibfnamefont {N.}~\bibnamefont
  {K{\"{a}}llman}}, \bibinfo {author} {\bibfnamefont {L.}~\bibnamefont
  {Grabitz}},\ and\ \bibinfo {author} {\bibfnamefont {P.}~\bibnamefont
  {T{\"{o}}rm{\"{a}}}},\ }\bibfield  {title} {\bibinfo {title}
  {{Pseudospin-orbit coupling and non-Hermitian effects in the quantum
  geometric tensor of a plasmonic lattice}},\ }\href
  {https://doi.org/10.1103/PhysRevB.109.165439} {\bibfield  {journal} {\bibinfo
   {journal} {Physical Review B}\ }\textbf {\bibinfo {volume} {109}},\ \bibinfo
  {pages} {165439} (\bibinfo {year} {2024}{\natexlab{a}})}\BibitemShut
  {NoStop}%
\bibitem [{\citenamefont {Liao}\ \emph {et~al.}(2021)\citenamefont {Liao},
  \citenamefont {Leblanc}, \citenamefont {Ren}, \citenamefont {Li},
  \citenamefont {Li}, \citenamefont {Solnyshkov}, \citenamefont {Malpuech},
  \citenamefont {Yao},\ and\ \citenamefont {Fu}}]{Liao:2021dd}%
  \BibitemOpen
  \bibfield  {author} {\bibinfo {author} {\bibfnamefont {Q.}~\bibnamefont
  {Liao}}, \bibinfo {author} {\bibfnamefont {C.}~\bibnamefont {Leblanc}},
  \bibinfo {author} {\bibfnamefont {J.}~\bibnamefont {Ren}}, \bibinfo {author}
  {\bibfnamefont {F.}~\bibnamefont {Li}}, \bibinfo {author} {\bibfnamefont
  {Y.}~\bibnamefont {Li}}, \bibinfo {author} {\bibfnamefont {D.}~\bibnamefont
  {Solnyshkov}}, \bibinfo {author} {\bibfnamefont {G.}~\bibnamefont
  {Malpuech}}, \bibinfo {author} {\bibfnamefont {J.}~\bibnamefont {Yao}},\ and\
  \bibinfo {author} {\bibfnamefont {H.}~\bibnamefont {Fu}},\ }\bibfield
  {title} {\bibinfo {title} {{Experimental Measurement of the Divergent Quantum
  Metric of an Exceptional Point}},\ }\href
  {https://doi.org/10.1103/PhysRevLett.127.107402} {\bibfield  {journal}
  {\bibinfo  {journal} {Physical Review Letters}\ }\textbf {\bibinfo {volume}
  {127}},\ \bibinfo {pages} {107402} (\bibinfo {year} {2021})}\BibitemShut
  {NoStop}%
\bibitem [{\citenamefont {Cuerda}\ \emph
  {et~al.}(2024{\natexlab{b}})\citenamefont {Cuerda}, \citenamefont {Taskinen},
  \citenamefont {K{\"{a}}llman}, \citenamefont {Grabitz},\ and\ \citenamefont
  {T{\"{o}}rm{\"{a}}}}]{Cuerda:2024jr}%
  \BibitemOpen
  \bibfield  {author} {\bibinfo {author} {\bibfnamefont {J.}~\bibnamefont
  {Cuerda}}, \bibinfo {author} {\bibfnamefont {J.~M.}\ \bibnamefont
  {Taskinen}}, \bibinfo {author} {\bibfnamefont {N.}~\bibnamefont
  {K{\"{a}}llman}}, \bibinfo {author} {\bibfnamefont {L.}~\bibnamefont
  {Grabitz}},\ and\ \bibinfo {author} {\bibfnamefont {P.}~\bibnamefont
  {T{\"{o}}rm{\"{a}}}},\ }\bibfield  {title} {\bibinfo {title} {{Observation of
  quantum metric and non-Hermitian Berry curvature in a plasmonic lattice}},\
  }\href {https://doi.org/10.1103/PhysRevResearch.6.L022020} {\bibfield
  {journal} {\bibinfo  {journal} {Physical Review Research}\ }\textbf {\bibinfo
  {volume} {6}},\ \bibinfo {pages} {L022020} (\bibinfo {year}
  {2024}{\natexlab{b}})}\BibitemShut {NoStop}%
\bibitem [{sup()}]{supplemental}%
  \BibitemOpen
  \href@noop {} {}\bibinfo {note} {The Supplemental Material has further
  details on the distance of eigenstates, real-space localization of wave
  packets, and electric susceptibility.}\BibitemShut {Stop}%
\bibitem [{\citenamefont {Silberstein}\ \emph {et~al.}(2020)\citenamefont
  {Silberstein}, \citenamefont {Behrends}, \citenamefont {Goldstein},\ and\
  \citenamefont {Ilan}}]{Silberstein:2020hi}%
  \BibitemOpen
  \bibfield  {author} {\bibinfo {author} {\bibfnamefont {N.}~\bibnamefont
  {Silberstein}}, \bibinfo {author} {\bibfnamefont {J.}~\bibnamefont
  {Behrends}}, \bibinfo {author} {\bibfnamefont {M.}~\bibnamefont
  {Goldstein}},\ and\ \bibinfo {author} {\bibfnamefont {R.}~\bibnamefont
  {Ilan}},\ }\bibfield  {title} {\bibinfo {title} {{Berry connection induced
  anomalous wave-packet dynamics in non-Hermitian systems}},\ }\href
  {https://doi.org/10.1103/PhysRevB.102.245147} {\bibfield  {journal} {\bibinfo
   {journal} {Physical Review B}\ }\textbf {\bibinfo {volume} {102}},\ \bibinfo
  {pages} {245147} (\bibinfo {year} {2020})}\BibitemShut {NoStop}%
\bibitem [{\citenamefont {Xu}\ \emph {et~al.}(2017)\citenamefont {Xu},
  \citenamefont {Wang},\ and\ \citenamefont {Duan}}]{Xu:2017bl}%
  \BibitemOpen
  \bibfield  {author} {\bibinfo {author} {\bibfnamefont {Y.}~\bibnamefont
  {Xu}}, \bibinfo {author} {\bibfnamefont {S.-T.}\ \bibnamefont {Wang}},\ and\
  \bibinfo {author} {\bibfnamefont {L.-M.}\ \bibnamefont {Duan}},\ }\bibfield
  {title} {\bibinfo {title} {{Weyl Exceptional Rings in a Three-Dimensional
  Dissipative Cold Atomic Gas}},\ }\href
  {https://doi.org/10.1103/PhysRevLett.118.045701} {\bibfield  {journal}
  {\bibinfo  {journal} {Physical Review Letters}\ }\textbf {\bibinfo {volume}
  {118}},\ \bibinfo {pages} {045701} (\bibinfo {year} {2017})}\BibitemShut
  {NoStop}%
\bibitem [{\citenamefont {Hu}\ \emph {et~al.}()\citenamefont {Hu},
  \citenamefont {Ostrovskaya},\ and\ \citenamefont {Estrecho}}]{hu2024role}%
  \BibitemOpen
  \bibfield  {author} {\bibinfo {author} {\bibfnamefont {Y.~M.~R.}\
  \bibnamefont {Hu}}, \bibinfo {author} {\bibfnamefont {E.~A.}\ \bibnamefont
  {Ostrovskaya}},\ and\ \bibinfo {author} {\bibfnamefont {E.}~\bibnamefont
  {Estrecho}},\ }\bibfield  {title} {\bibinfo {title} {{Role of quantum
  geometric tensor on wavepacket dynamics in two-dimensional non-Hermitian
  systems}},\ }\Eprint {https://arxiv.org/abs/2412.08141} {arXiv:2412.08141}
  \BibitemShut {NoStop}%
\bibitem [{\citenamefont {Singhal}\ \emph {et~al.}(2023)\citenamefont
  {Singhal}, \citenamefont {Martello}, \citenamefont {Agrawal}, \citenamefont
  {Ozawa}, \citenamefont {Price},\ and\ \citenamefont
  {Gadway}}]{Singhal:2023gl}%
  \BibitemOpen
  \bibfield  {author} {\bibinfo {author} {\bibfnamefont {Y.}~\bibnamefont
  {Singhal}}, \bibinfo {author} {\bibfnamefont {E.}~\bibnamefont {Martello}},
  \bibinfo {author} {\bibfnamefont {S.}~\bibnamefont {Agrawal}}, \bibinfo
  {author} {\bibfnamefont {T.}~\bibnamefont {Ozawa}}, \bibinfo {author}
  {\bibfnamefont {H.}~\bibnamefont {Price}},\ and\ \bibinfo {author}
  {\bibfnamefont {B.}~\bibnamefont {Gadway}},\ }\bibfield  {title} {\bibinfo
  {title} {{Measuring the adiabatic non-Hermitian Berry phase in
  feedback-coupled oscillators}},\ }\href
  {https://doi.org/10.1103/PhysRevResearch.5.L032026} {\bibfield  {journal}
  {\bibinfo  {journal} {Physical Review Research}\ }\textbf {\bibinfo {volume}
  {5}},\ \bibinfo {pages} {L032026} (\bibinfo {year} {2023})}\BibitemShut
  {NoStop}%
\bibitem [{\citenamefont {Ozawa}\ and\ \citenamefont
  {Schomerus}()}]{Ozawa2024}%
  \BibitemOpen
  \bibfield  {author} {\bibinfo {author} {\bibfnamefont {T.}~\bibnamefont
  {Ozawa}}\ and\ \bibinfo {author} {\bibfnamefont {H.}~\bibnamefont
  {Schomerus}},\ }\bibfield  {title} {\bibinfo {title} {{Geometric contribution
  to adiabatic amplification in non-Hermitian systems}},\ }\Eprint
  {https://arxiv.org/abs/2409.13595} {arXiv:2409.13595} \BibitemShut {NoStop}%
\bibitem [{\citenamefont {Callen}\ and\ \citenamefont
  {Welton}(1951)}]{Callen:1951be}%
  \BibitemOpen
  \bibfield  {author} {\bibinfo {author} {\bibfnamefont {H.~B.}\ \bibnamefont
  {Callen}}\ and\ \bibinfo {author} {\bibfnamefont {T.~A.}\ \bibnamefont
  {Welton}},\ }\bibfield  {title} {\bibinfo {title} {{Irreversibility and
  Generalized Noise}},\ }\href {https://doi.org/10.1103/PhysRev.83.34}
  {\bibfield  {journal} {\bibinfo  {journal} {Physical Review}\ }\textbf
  {\bibinfo {volume} {83}},\ \bibinfo {pages} {34} (\bibinfo {year}
  {1951})}\BibitemShut {NoStop}%
\bibitem [{\citenamefont {Kubo}(1957)}]{Kubo:1957cl}%
  \BibitemOpen
  \bibfield  {author} {\bibinfo {author} {\bibfnamefont {R.}~\bibnamefont
  {Kubo}},\ }\bibfield  {title} {\bibinfo {title} {{Statistical-Mechanical
  Theory of Irreversible Processes. I. General Theory and Simple Applications
  to Magnetic and Conduction Problems}},\ }\href
  {https://doi.org/10.1143/JPSJ.12.570} {\bibfield  {journal} {\bibinfo
  {journal} {Journal of the Physical Society of Japan}\ }\textbf {\bibinfo
  {volume} {12}},\ \bibinfo {pages} {570} (\bibinfo {year} {1957})}\BibitemShut
  {NoStop}%
\bibitem [{\citenamefont {Kubo}(1966)}]{Kubo:1966dq}%
  \BibitemOpen
  \bibfield  {author} {\bibinfo {author} {\bibfnamefont {R.}~\bibnamefont
  {Kubo}},\ }\bibfield  {title} {\bibinfo {title} {{The fluctuation-dissipation
  theorem}},\ }\href {https://doi.org/10.1088/0034-4885/29/1/306} {\bibfield
  {journal} {\bibinfo  {journal} {Reports on Progress in Physics}\ }\textbf
  {\bibinfo {volume} {29}},\ \bibinfo {pages} {306} (\bibinfo {year}
  {1966})}\BibitemShut {NoStop}%
\bibitem [{\citenamefont {Sun}\ and\ \citenamefont {Zheng}(2019)}]{Sun:2019en}%
  \BibitemOpen
  \bibfield  {author} {\bibinfo {author} {\bibfnamefont {S.}~\bibnamefont
  {Sun}}\ and\ \bibinfo {author} {\bibfnamefont {Y.}~\bibnamefont {Zheng}},\
  }\bibfield  {title} {\bibinfo {title} {{Distinct Bound of the Quantum Speed
  Limit via the Gauge Invariant Distance}},\ }\href {https://doi.org/2}
  {\bibfield  {journal} {\bibinfo  {journal} {Physical Review Letters}\
  }\textbf {\bibinfo {volume} {123}},\ \bibinfo {pages} {180403} (\bibinfo
  {year} {2019})}\BibitemShut {NoStop}%
\bibitem [{\citenamefont {Braunstein}\ and\ \citenamefont
  {Caves}(1994)}]{Braunstein:1994jl}%
  \BibitemOpen
  \bibfield  {author} {\bibinfo {author} {\bibfnamefont {S.~L.}\ \bibnamefont
  {Braunstein}}\ and\ \bibinfo {author} {\bibfnamefont {C.~M.}\ \bibnamefont
  {Caves}},\ }\bibfield  {title} {\bibinfo {title} {{Statistical distance and
  the geometry of quantum states}},\ }\href
  {https://doi.org/10.1103/PhysRevLett.72.3439} {\bibfield  {journal} {\bibinfo
   {journal} {Physical Review Letters}\ }\textbf {\bibinfo {volume} {72}},\
  \bibinfo {pages} {3439} (\bibinfo {year} {1994})}\BibitemShut {NoStop}%
\bibitem [{\citenamefont {Cui}\ and\ \citenamefont {Zheng}(2012)}]{Cui:2012kg}%
  \BibitemOpen
  \bibfield  {author} {\bibinfo {author} {\bibfnamefont {X.-D.}\ \bibnamefont
  {Cui}}\ and\ \bibinfo {author} {\bibfnamefont {Y.}~\bibnamefont {Zheng}},\
  }\bibfield  {title} {\bibinfo {title} {{Geometric phases in non-Hermitian
  quantum mechanics}},\ }\href {https://doi.org/10.1103/PhysRevA.86.064104}
  {\bibfield  {journal} {\bibinfo  {journal} {Physical Review A}\ }\textbf
  {\bibinfo {volume} {86}},\ \bibinfo {pages} {064104} (\bibinfo {year}
  {2012})}\BibitemShut {NoStop}%
\bibitem [{\citenamefont {Brody}\ and\ \citenamefont
  {Graefe}(2013)}]{Brody:2013et}%
  \BibitemOpen
  \bibfield  {author} {\bibinfo {author} {\bibfnamefont {D.}~\bibnamefont
  {Brody}}\ and\ \bibinfo {author} {\bibfnamefont {E.-M.}\ \bibnamefont
  {Graefe}},\ }\bibfield  {title} {\bibinfo {title} {{Information Geometry of
  Complex Hamiltonians and Exceptional Points}},\ }\href
  {https://doi.org/10.3390/e15093361} {\bibfield  {journal} {\bibinfo
  {journal} {Entropy}\ }\textbf {\bibinfo {volume} {15}},\ \bibinfo {pages}
  {3361} (\bibinfo {year} {2013})}\BibitemShut {NoStop}%
\bibitem [{\citenamefont {Hu}\ \emph {et~al.}(2024)\citenamefont {Hu},
  \citenamefont {Ostrovskaya},\ and\ \citenamefont {Estrecho}}]{Hu:2024kw}%
  \BibitemOpen
  \bibfield  {author} {\bibinfo {author} {\bibfnamefont {Y.-M.~R.}\
  \bibnamefont {Hu}}, \bibinfo {author} {\bibfnamefont {E.~A.}\ \bibnamefont
  {Ostrovskaya}},\ and\ \bibinfo {author} {\bibfnamefont {E.}~\bibnamefont
  {Estrecho}},\ }\bibfield  {title} {\bibinfo {title} {{Generalized quantum
  geometric tensor in a non-Hermitian exciton-polariton system [Invited]}},\
  }\href {https://doi.org/10.1364/OME.497010} {\bibfield  {journal} {\bibinfo
  {journal} {Optical Materials Express}\ }\textbf {\bibinfo {volume} {14}},\
  \bibinfo {pages} {664} (\bibinfo {year} {2024})}\BibitemShut {NoStop}%
\bibitem [{\citenamefont {{Chen Ye}}\ \emph {et~al.}(2024)\citenamefont {{Chen
  Ye}}, \citenamefont {Vleeshouwers}, \citenamefont {Heatley}, \citenamefont
  {Gritsev},\ and\ \citenamefont {{Morais Smith}}}]{Chen:2024ij}%
  \BibitemOpen
  \bibfield  {author} {\bibinfo {author} {\bibfnamefont {C.}~\bibnamefont
  {{Chen Ye}}}, \bibinfo {author} {\bibfnamefont {W.~L.}\ \bibnamefont
  {Vleeshouwers}}, \bibinfo {author} {\bibfnamefont {S.}~\bibnamefont
  {Heatley}}, \bibinfo {author} {\bibfnamefont {V.}~\bibnamefont {Gritsev}},\
  and\ \bibinfo {author} {\bibfnamefont {C.}~\bibnamefont {{Morais Smith}}},\
  }\bibfield  {title} {\bibinfo {title} {{Quantum metric of non-Hermitian
  Su-Schrieffer-Heeger systems}},\ }\href
  {https://doi.org/10.1103/PhysRevResearch.6.023202} {\bibfield  {journal}
  {\bibinfo  {journal} {Physical Review Research}\ }\textbf {\bibinfo {volume}
  {6}},\ \bibinfo {pages} {023202} (\bibinfo {year} {2024})}\BibitemShut
  {NoStop}%
\bibitem [{\citenamefont {Orlov}\ \emph {et~al.}()\citenamefont {Orlov},
  \citenamefont {Shlyapnikov},\ and\ \citenamefont
  {Kurlov}}]{orlov2024adiabatic}%
  \BibitemOpen
  \bibfield  {author} {\bibinfo {author} {\bibfnamefont {P.}~\bibnamefont
  {Orlov}}, \bibinfo {author} {\bibfnamefont {G.~V.}\ \bibnamefont
  {Shlyapnikov}},\ and\ \bibinfo {author} {\bibfnamefont {D.~V.}\ \bibnamefont
  {Kurlov}},\ }\href {https://arxiv.org/abs/2404.12337} {\bibinfo {title}
  {Adiabatic transformations in dissipative and non-hermitian phase
  transitions}},\ \Eprint {https://arxiv.org/abs/2404.12337} {arXiv:2404.12337}
  \BibitemShut {NoStop}%
\bibitem [{\citenamefont {Zhu}\ \emph {et~al.}(2021)\citenamefont {Zhu},
  \citenamefont {Zheng}, \citenamefont {Zhu},\ and\ \citenamefont
  {Palumbo}}]{Zhu:2021gy}%
  \BibitemOpen
  \bibfield  {author} {\bibinfo {author} {\bibfnamefont {Y.-Q.}\ \bibnamefont
  {Zhu}}, \bibinfo {author} {\bibfnamefont {W.}~\bibnamefont {Zheng}}, \bibinfo
  {author} {\bibfnamefont {S.-L.}\ \bibnamefont {Zhu}},\ and\ \bibinfo {author}
  {\bibfnamefont {G.}~\bibnamefont {Palumbo}},\ }\bibfield  {title} {\bibinfo
  {title} {{Band topology of pseudo-Hermitian phases through tensor Berry
  connections and quantum metric}},\ }\href
  {https://doi.org/10.1103/PhysRevB.104.205103} {\bibfield  {journal} {\bibinfo
   {journal} {Physical Review B}\ }\textbf {\bibinfo {volume} {104}},\ \bibinfo
  {pages} {205103} (\bibinfo {year} {2021})}\BibitemShut {NoStop}%
\bibitem [{\citenamefont {Cai}\ \emph {et~al.}(2019)\citenamefont {Cai},
  \citenamefont {Meng}, \citenamefont {Zhang},\ and\ \citenamefont
  {Wang}}]{Cai:2019hc}%
  \BibitemOpen
  \bibfield  {author} {\bibinfo {author} {\bibfnamefont {X.}~\bibnamefont
  {Cai}}, \bibinfo {author} {\bibfnamefont {R.}~\bibnamefont {Meng}}, \bibinfo
  {author} {\bibfnamefont {Y.}~\bibnamefont {Zhang}},\ and\ \bibinfo {author}
  {\bibfnamefont {L.}~\bibnamefont {Wang}},\ }\bibfield  {title} {\bibinfo
  {title} {{Geometry of quantum evolution in a nonequilibrium environment}},\
  }\href {https://doi.org/10.1209/0295-5075/125/30007} {\bibfield  {journal}
  {\bibinfo  {journal} {EPL (Europhysics Letters)}\ }\textbf {\bibinfo {volume}
  {125}},\ \bibinfo {pages} {30007} (\bibinfo {year} {2019})}\BibitemShut
  {NoStop}%
\bibitem [{\citenamefont {H{\"{o}}rnedal}\ \emph {et~al.}()\citenamefont
  {H{\"{o}}rnedal}, \citenamefont {Pro{\'{s}}niak}, \citenamefont {del Campo},\
  and\ \citenamefont {Chenu}}]{Hornedal2024}%
  \BibitemOpen
  \bibfield  {author} {\bibinfo {author} {\bibfnamefont {N.}~\bibnamefont
  {H{\"{o}}rnedal}}, \bibinfo {author} {\bibfnamefont {O.~A.}\ \bibnamefont
  {Pro{\'{s}}niak}}, \bibinfo {author} {\bibfnamefont {A.}~\bibnamefont {del
  Campo}},\ and\ \bibinfo {author} {\bibfnamefont {A.}~\bibnamefont {Chenu}},\
  }\bibfield  {title} {\bibinfo {title} {{A geometrical description of
  non-Hermitian dynamics: speed limits in finite rank density operators}},\
  }\Eprint {https://arxiv.org/abs/2405.13913} {arXiv:2405.13913} \BibitemShut
  {NoStop}%
\bibitem [{\citenamefont {Meden}\ \emph {et~al.}(2023)\citenamefont {Meden},
  \citenamefont {Grunwald},\ and\ \citenamefont {Kennes}}]{Meden:2023ju}%
  \BibitemOpen
  \bibfield  {author} {\bibinfo {author} {\bibfnamefont {V.}~\bibnamefont
  {Meden}}, \bibinfo {author} {\bibfnamefont {L.}~\bibnamefont {Grunwald}},\
  and\ \bibinfo {author} {\bibfnamefont {D.~M.}\ \bibnamefont {Kennes}},\
  }\bibfield  {title} {\bibinfo {title} {{$\mathcal{PT}$-symmetric,
  non-Hermitian quantum many-body physics—a methodological perspective}},\
  }\href {https://doi.org/10.1088/1361-6633/ad05f3} {\bibfield  {journal}
  {\bibinfo  {journal} {Reports on Progress in Physics}\ }\textbf {\bibinfo
  {volume} {86}},\ \bibinfo {pages} {124501} (\bibinfo {year}
  {2023})}\BibitemShut {NoStop}%
\bibitem [{\citenamefont {Bandyopadhyay}\ \emph {et~al.}()\citenamefont
  {Bandyopadhyay}, \citenamefont {Hauke},\ and\ \citenamefont
  {Roy}}]{Bandyopadhyay2024}%
  \BibitemOpen
  \bibfield  {author} {\bibinfo {author} {\bibfnamefont {S.}~\bibnamefont
  {Bandyopadhyay}}, \bibinfo {author} {\bibfnamefont {P.}~\bibnamefont
  {Hauke}},\ and\ \bibinfo {author} {\bibfnamefont {S.~S.}\ \bibnamefont
  {Roy}},\ }\bibfield  {title} {\bibinfo {title} {{Quantifying non-Hermiticity
  using single- and many-particle quantum properties}},\ }\Eprint
  {https://arxiv.org/abs/2406.13517} {arXiv:2406.13517} \BibitemShut {NoStop}%
\bibitem [{\citenamefont {Brody}(2014)}]{Brody:2014jv}%
  \BibitemOpen
  \bibfield  {author} {\bibinfo {author} {\bibfnamefont {D.~C.}\ \bibnamefont
  {Brody}},\ }\bibfield  {title} {\bibinfo {title} {{Biorthogonal quantum
  mechanics}},\ }\href {https://doi.org/10.1088/1751-8113/47/3/035305}
  {\bibfield  {journal} {\bibinfo  {journal} {Journal of Physics A:
  Mathematical and Theoretical}\ }\textbf {\bibinfo {volume} {47}},\ \bibinfo
  {pages} {035305} (\bibinfo {year} {2014})}\BibitemShut {NoStop}%
\bibitem [{\citenamefont {Frith}(2019)}]{Frith2019}%
  \BibitemOpen
  \bibfield  {author} {\bibinfo {author} {\bibfnamefont {T.~D.}\ \bibnamefont
  {Frith}},\ }\emph {\bibinfo {title} {Time-dependence in non-Hermitian quantum
  systems}},\ \href {https://openaccess.city.ac.uk/id/eprint/23897/} {Ph.D.
  thesis},\ \bibinfo  {school} {City, University of London} (\bibinfo {year}
  {2019})\BibitemShut {NoStop}%
\bibitem [{\citenamefont {Bender}\ and\ \citenamefont
  {Boettcher}(1998)}]{Bender:1998bw}%
  \BibitemOpen
  \bibfield  {author} {\bibinfo {author} {\bibfnamefont {C.~M.}\ \bibnamefont
  {Bender}}\ and\ \bibinfo {author} {\bibfnamefont {S.}~\bibnamefont
  {Boettcher}},\ }\bibfield  {title} {\bibinfo {title} {{Real Spectra in
  Non-Hermitian Hamiltonians Having $\mathcal{P}\mathcal{T}$ Symmetry}},\
  }\href {https://doi.org/10.1103/PhysRevLett.80.5243} {\bibfield  {journal}
  {\bibinfo  {journal} {Physical Review Letters}\ }\textbf {\bibinfo {volume}
  {80}},\ \bibinfo {pages} {5243} (\bibinfo {year} {1998})}\BibitemShut
  {NoStop}%
\bibitem [{\citenamefont {Bender}(2005)}]{Bender:2005cd}%
  \BibitemOpen
  \bibfield  {author} {\bibinfo {author} {\bibfnamefont {C.~M.}\ \bibnamefont
  {Bender}},\ }\bibfield  {title} {\bibinfo {title} {{Introduction to
  $\mathcal{P}\mathcal{T}$-symmetric quantum theory}},\ }\href
  {https://doi.org/10.1080/00107500072632} {\bibfield  {journal} {\bibinfo
  {journal} {Contemporary Physics}\ }\textbf {\bibinfo {volume} {46}},\
  \bibinfo {pages} {277} (\bibinfo {year} {2005})}\BibitemShut {NoStop}%
\bibitem [{Note2()}]{Note2}%
  \BibitemOpen
  \bibinfo {note} {In $d$-dimensional systems, we use either $\DOTSI \intop
  \ilimits@ _\protect \mathbf {k} = L^{-d} \DOTSB \sum@ \slimits@ _\protect
  \mathbf {k}$ for periodic boundary conditions with linear dimension $L$, or
  $\DOTSI \intop \ilimits@ _\protect \mathbf {k} = (2\pi )^{-d} \DOTSI \intop
  \ilimits@ d^d \protect \mathbf {k}$ for an infinite system.}\BibitemShut
  {Stop}%
\bibitem [{\citenamefont {Souza}\ \emph {et~al.}(2000)\citenamefont {Souza},
  \citenamefont {Wilkens},\ and\ \citenamefont {Martin}}]{Souza:2000cj}%
  \BibitemOpen
  \bibfield  {author} {\bibinfo {author} {\bibfnamefont {I.}~\bibnamefont
  {Souza}}, \bibinfo {author} {\bibfnamefont {T.}~\bibnamefont {Wilkens}},\
  and\ \bibinfo {author} {\bibfnamefont {R.~M.}\ \bibnamefont {Martin}},\
  }\bibfield  {title} {\bibinfo {title} {{Polarization and localization in
  insulators: Generating function approach}},\ }\href
  {https://doi.org/10.1103/PhysRevB.62.1666} {\bibfield  {journal} {\bibinfo
  {journal} {Physical Review B}\ }\textbf {\bibinfo {volume} {62}},\ \bibinfo
  {pages} {1666} (\bibinfo {year} {2000})}\BibitemShut {NoStop}%
\bibitem [{\citenamefont {Sticlet}\ \emph {et~al.}(2022)\citenamefont
  {Sticlet}, \citenamefont {D{\'{o}}ra},\ and\ \citenamefont
  {Moca}}]{Sticlet:2022en}%
  \BibitemOpen
  \bibfield  {author} {\bibinfo {author} {\bibfnamefont {D.}~\bibnamefont
  {Sticlet}}, \bibinfo {author} {\bibfnamefont {B.}~\bibnamefont
  {D{\'{o}}ra}},\ and\ \bibinfo {author} {\bibfnamefont {C.~P.}\ \bibnamefont
  {Moca}},\ }\bibfield  {title} {\bibinfo {title} {{Kubo Formula for
  Non-Hermitian Systems and Tachyon Optical Conductivity}},\ }\href
  {https://doi.org/10.1103/PhysRevLett.128.016802} {\bibfield  {journal}
  {\bibinfo  {journal} {Physical Review Letters}\ }\textbf {\bibinfo {volume}
  {128}},\ \bibinfo {pages} {016802} (\bibinfo {year} {2022})}\BibitemShut
  {NoStop}%
\bibitem [{\citenamefont {Pan}\ \emph {et~al.}(2020)\citenamefont {Pan},
  \citenamefont {Chen}, \citenamefont {Chen},\ and\ \citenamefont
  {Zhai}}]{Pan:2020hp}%
  \BibitemOpen
  \bibfield  {author} {\bibinfo {author} {\bibfnamefont {L.}~\bibnamefont
  {Pan}}, \bibinfo {author} {\bibfnamefont {X.}~\bibnamefont {Chen}}, \bibinfo
  {author} {\bibfnamefont {Y.}~\bibnamefont {Chen}},\ and\ \bibinfo {author}
  {\bibfnamefont {H.}~\bibnamefont {Zhai}},\ }\bibfield  {title} {\bibinfo
  {title} {{Non-Hermitian linear response theory}},\ }\href
  {https://doi.org/10.1038/s41567-020-0889-6} {\bibfield  {journal} {\bibinfo
  {journal} {Nature Physics}\ }\textbf {\bibinfo {volume} {16}},\ \bibinfo
  {pages} {767} (\bibinfo {year} {2020})}\BibitemShut {NoStop}%
\bibitem [{\citenamefont {Griffiths}(2017)}]{Griffiths:2017jp}%
  \BibitemOpen
  \bibfield  {author} {\bibinfo {author} {\bibfnamefont {D.~J.}\ \bibnamefont
  {Griffiths}},\ }\href {https://doi.org/10.1017/9781108333511} {\emph
  {\bibinfo {title} {Introduction to Electrodynamics}}},\ \bibinfo {edition}
  {4th}\ ed.\ (\bibinfo  {publisher} {Cambridge University Press},\ \bibinfo
  {address} {Cambridge, U.K.},\ \bibinfo {year} {2017})\BibitemShut {NoStop}%
\bibitem [{\citenamefont {King-Smith}\ and\ \citenamefont
  {Vanderbilt}(1993)}]{KingSmith:1993fn}%
  \BibitemOpen
  \bibfield  {author} {\bibinfo {author} {\bibfnamefont {R.~D.}\ \bibnamefont
  {King-Smith}}\ and\ \bibinfo {author} {\bibfnamefont {D.}~\bibnamefont
  {Vanderbilt}},\ }\bibfield  {title} {\bibinfo {title} {{Theory of
  polarization of crystalline solids}},\ }\href
  {https://doi.org/10.1103/PhysRevB.47.1651} {\bibfield  {journal} {\bibinfo
  {journal} {Physical Review B}\ }\textbf {\bibinfo {volume} {47}},\ \bibinfo
  {pages} {1651} (\bibinfo {year} {1993})}\BibitemShut {NoStop}%
\bibitem [{\citenamefont {Vanderbilt}\ and\ \citenamefont
  {King-Smith}(1993)}]{Vanderbilt:1993dx}%
  \BibitemOpen
  \bibfield  {author} {\bibinfo {author} {\bibfnamefont {D.}~\bibnamefont
  {Vanderbilt}}\ and\ \bibinfo {author} {\bibfnamefont {R.~D.}\ \bibnamefont
  {King-Smith}},\ }\bibfield  {title} {\bibinfo {title} {{Electric polarization
  as a bulk quantity and its relation to surface charge}},\ }\href
  {https://doi.org/10.1103/PhysRevB.48.4442} {\bibfield  {journal} {\bibinfo
  {journal} {Physical Review B}\ }\textbf {\bibinfo {volume} {48}},\ \bibinfo
  {pages} {4442} (\bibinfo {year} {1993})}\BibitemShut {NoStop}%
\bibitem [{\citenamefont {Scalapino}\ \emph {et~al.}(1993)\citenamefont
  {Scalapino}, \citenamefont {White},\ and\ \citenamefont
  {Zhang}}]{Scalapino:1993ih}%
  \BibitemOpen
  \bibfield  {author} {\bibinfo {author} {\bibfnamefont {D.~J.}\ \bibnamefont
  {Scalapino}}, \bibinfo {author} {\bibfnamefont {S.~R.}\ \bibnamefont
  {White}},\ and\ \bibinfo {author} {\bibfnamefont {S.}~\bibnamefont {Zhang}},\
  }\bibfield  {title} {\bibinfo {title} {{Insulator, metal, or superconductor:
  The criteria}},\ }\href {https://doi.org/10.1103/PhysRevB.47.7995} {\bibfield
   {journal} {\bibinfo  {journal} {Physical Review B}\ }\textbf {\bibinfo
  {volume} {47}},\ \bibinfo {pages} {7995} (\bibinfo {year}
  {1993})}\BibitemShut {NoStop}%
\bibitem [{\citenamefont {Altland}\ and\ \citenamefont
  {Simons}(2010)}]{AltlandSimons}%
  \BibitemOpen
  \bibfield  {author} {\bibinfo {author} {\bibfnamefont {A.}~\bibnamefont
  {Altland}}\ and\ \bibinfo {author} {\bibfnamefont {B.~D.}\ \bibnamefont
  {Simons}},\ }\href {https://doi.org/10.1017/CBO9780511789984} {\emph
  {\bibinfo {title} {{Condensed Matter Field Theory}}}},\ \bibinfo {edition}
  {2nd}\ ed.\ (\bibinfo  {publisher} {Cambridge University Press},\ \bibinfo
  {address} {Cambridge, U.K.},\ \bibinfo {year} {2010})\BibitemShut {NoStop}%
\bibitem [{\citenamefont {Herviou}\ \emph {et~al.}(2019)\citenamefont
  {Herviou}, \citenamefont {Regnault},\ and\ \citenamefont
  {Bardarson}}]{Herviou:2019ku}%
  \BibitemOpen
  \bibfield  {author} {\bibinfo {author} {\bibfnamefont {L.}~\bibnamefont
  {Herviou}}, \bibinfo {author} {\bibfnamefont {N.}~\bibnamefont {Regnault}},\
  and\ \bibinfo {author} {\bibfnamefont {J.~H.}\ \bibnamefont {Bardarson}},\
  }\bibfield  {title} {\bibinfo {title} {{Entanglement spectrum and symmetries
  in non-Hermitian fermionic non-interacting models}},\ }\href
  {https://doi.org/10.21468/SciPostPhys.7.5.069} {\bibfield  {journal}
  {\bibinfo  {journal} {SciPost Physics}\ }\textbf {\bibinfo {volume} {7}},\
  \bibinfo {pages} {069} (\bibinfo {year} {2019})}\BibitemShut {NoStop}%
\bibitem [{\citenamefont {Lee}\ \emph {et~al.}(2020)\citenamefont {Lee},
  \citenamefont {Lee},\ and\ \citenamefont {Yang}}]{Lee:2020ex}%
  \BibitemOpen
  \bibfield  {author} {\bibinfo {author} {\bibfnamefont {E.}~\bibnamefont
  {Lee}}, \bibinfo {author} {\bibfnamefont {H.}~\bibnamefont {Lee}},\ and\
  \bibinfo {author} {\bibfnamefont {B.-J.}\ \bibnamefont {Yang}},\ }\bibfield
  {title} {\bibinfo {title} {{Many-body approach to non-Hermitian physics in
  fermionic systems}},\ }\href {https://doi.org/10.1103/PhysRevB.101.121109}
  {\bibfield  {journal} {\bibinfo  {journal} {Physical Review B}\ }\textbf
  {\bibinfo {volume} {101}},\ \bibinfo {pages} {121109} (\bibinfo {year}
  {2020})}\BibitemShut {NoStop}%
\bibitem [{\citenamefont {Alsallom}\ \emph {et~al.}(2022)\citenamefont
  {Alsallom}, \citenamefont {Herviou}, \citenamefont {Yazyev},\ and\
  \citenamefont {Brzezi{\'{n}}ska}}]{Alsallom:2022jz}%
  \BibitemOpen
  \bibfield  {author} {\bibinfo {author} {\bibfnamefont {F.}~\bibnamefont
  {Alsallom}}, \bibinfo {author} {\bibfnamefont {L.}~\bibnamefont {Herviou}},
  \bibinfo {author} {\bibfnamefont {O.~V.}\ \bibnamefont {Yazyev}},\ and\
  \bibinfo {author} {\bibfnamefont {M.}~\bibnamefont {Brzezi{\'{n}}ska}},\
  }\bibfield  {title} {\bibinfo {title} {{Fate of the non-Hermitian skin effect
  in many-body fermionic systems}},\ }\href
  {https://doi.org/10.1103/PhysRevResearch.4.033122} {\bibfield  {journal}
  {\bibinfo  {journal} {Physical Review Research}\ }\textbf {\bibinfo {volume}
  {4}},\ \bibinfo {pages} {70} (\bibinfo {year} {2022})}\BibitemShut {NoStop}%
\bibitem [{\citenamefont {Kos}\ and\ \citenamefont
  {Prosen}(2017)}]{Kos:2017ew}%
  \BibitemOpen
  \bibfield  {author} {\bibinfo {author} {\bibfnamefont {P.}~\bibnamefont
  {Kos}}\ and\ \bibinfo {author} {\bibfnamefont {T.}~\bibnamefont {Prosen}},\
  }\bibfield  {title} {\bibinfo {title} {{Time-dependent correlation functions
  in open quadratic fermionic systems}},\ }\href
  {https://doi.org/10.1088/1742-5468/aa9681} {\bibfield  {journal} {\bibinfo
  {journal} {Journal of Statistical Mechanics: Theory and Experiment}\ }\textbf
  {\bibinfo {volume} {2017}},\ \bibinfo {pages} {123103} (\bibinfo {year}
  {2017})}\BibitemShut {NoStop}%
\bibitem [{\citenamefont {Kunst}\ \emph {et~al.}(2018)\citenamefont {Kunst},
  \citenamefont {Edvardsson}, \citenamefont {Budich},\ and\ \citenamefont
  {Bergholtz}}]{Kunst:2018ku}%
  \BibitemOpen
  \bibfield  {author} {\bibinfo {author} {\bibfnamefont {F.~K.}\ \bibnamefont
  {Kunst}}, \bibinfo {author} {\bibfnamefont {E.}~\bibnamefont {Edvardsson}},
  \bibinfo {author} {\bibfnamefont {J.~C.}\ \bibnamefont {Budich}},\ and\
  \bibinfo {author} {\bibfnamefont {E.~J.}\ \bibnamefont {Bergholtz}},\
  }\bibfield  {title} {\bibinfo {title} {{Biorthogonal Bulk-Boundary
  Correspondence in Non-Hermitian Systems}},\ }\href
  {https://doi.org/10.1103/PhysRevLett.121.026808} {\bibfield  {journal}
  {\bibinfo  {journal} {Physical Review Letters}\ }\textbf {\bibinfo {volume}
  {121}},\ \bibinfo {pages} {026808} (\bibinfo {year} {2018})}\BibitemShut
  {NoStop}%
\bibitem [{\citenamefont {Kolodrubetz}\ \emph {et~al.}(2017)\citenamefont
  {Kolodrubetz}, \citenamefont {Sels}, \citenamefont {Mehta},\ and\
  \citenamefont {Polkovnikov}}]{Kolodrubetz:2017jg}%
  \BibitemOpen
  \bibfield  {author} {\bibinfo {author} {\bibfnamefont {M.}~\bibnamefont
  {Kolodrubetz}}, \bibinfo {author} {\bibfnamefont {D.}~\bibnamefont {Sels}},
  \bibinfo {author} {\bibfnamefont {P.}~\bibnamefont {Mehta}},\ and\ \bibinfo
  {author} {\bibfnamefont {A.}~\bibnamefont {Polkovnikov}},\ }\bibfield
  {title} {\bibinfo {title} {{Geometry and non-adiabatic response in quantum
  and classical systems}},\ }\href
  {https://doi.org/10.1016/j.physrep.2017.07.001} {\bibfield  {journal}
  {\bibinfo  {journal} {Physics Repports}\ }\textbf {\bibinfo {volume} {697}},\
  \bibinfo {pages} {1} (\bibinfo {year} {2017})}\BibitemShut {NoStop}%
\bibitem [{\citenamefont {Dalibard}\ \emph {et~al.}(2011)\citenamefont
  {Dalibard}, \citenamefont {Gerbier}, \citenamefont {Juzeliunas},\ and\
  \citenamefont {{\"{O}}hberg}}]{Dalibard:2011gg}%
  \BibitemOpen
  \bibfield  {author} {\bibinfo {author} {\bibfnamefont {J.}~\bibnamefont
  {Dalibard}}, \bibinfo {author} {\bibfnamefont {F.}~\bibnamefont {Gerbier}},
  \bibinfo {author} {\bibfnamefont {G.}~\bibnamefont {Juzeliunas}},\ and\
  \bibinfo {author} {\bibfnamefont {P.}~\bibnamefont {{\"{O}}hberg}},\
  }\bibfield  {title} {\bibinfo {title} {{Colloquium: Artificial gauge
  potentials for neutral atoms}},\ }\href
  {https://doi.org/10.1103/RevModPhys.83.1523} {\bibfield  {journal} {\bibinfo
  {journal} {Reviews of Modern Physics}\ }\textbf {\bibinfo {volume} {83}},\
  \bibinfo {pages} {1523} (\bibinfo {year} {2011})}\BibitemShut {NoStop}%
\bibitem [{\citenamefont {Resta}(1998)}]{Resta:1998hj}%
  \BibitemOpen
  \bibfield  {author} {\bibinfo {author} {\bibfnamefont {R.}~\bibnamefont
  {Resta}},\ }\bibfield  {title} {\bibinfo {title} {{Quantum-Mechanical
  Position Operator in Extended Systems}},\ }\href
  {https://doi.org/10.1103/PhysRevLett.80.1800} {\bibfield  {journal} {\bibinfo
   {journal} {Physical Review Letters}\ }\textbf {\bibinfo {volume} {80}},\
  \bibinfo {pages} {1800} (\bibinfo {year} {1998})}\BibitemShut {NoStop}%
\bibitem [{\citenamefont {Kolodrubetz}\ \emph {et~al.}(2013)\citenamefont
  {Kolodrubetz}, \citenamefont {Gritsev},\ and\ \citenamefont
  {Polkovnikov}}]{Kolodrubetz:2013bf}%
  \BibitemOpen
  \bibfield  {author} {\bibinfo {author} {\bibfnamefont {M.}~\bibnamefont
  {Kolodrubetz}}, \bibinfo {author} {\bibfnamefont {V.}~\bibnamefont
  {Gritsev}},\ and\ \bibinfo {author} {\bibfnamefont {A.}~\bibnamefont
  {Polkovnikov}},\ }\bibfield  {title} {\bibinfo {title} {{Classifying and
  measuring geometry of a quantum ground state manifold}},\ }\href
  {https://doi.org/10.1103/PhysRevB.88.064304} {\bibfield  {journal} {\bibinfo
  {journal} {Physical Review B}\ }\textbf {\bibinfo {volume} {88}},\ \bibinfo
  {pages} {064304} (\bibinfo {year} {2013})}\BibitemShut {NoStop}%
\end{thebibliography}%

\end{document}